\documentclass[aps,prb,twocolumn,superscriptaddress,showpacs]{revtex4}

\usepackage{graphics}
\usepackage{amsmath}
\usepackage{bm}

\begin{document}

\title{Bosonization of strongly interacting electrons}

\author{K. A. Matveev}

\affiliation{Materials Science Division, Argonne National Laboratory, 
Argonne, Illinois 60439, USA}

\author{A. Furusaki}

\affiliation{Condensed Matter Theory Laboratory, RIKEN, Wako, Saitama
  351-0198, Japan} 

\author{L. I. Glazman}

\affiliation{Theoretical Physics Institute, University of Minnesota, 
Minneapolis, Minnesota 55455, USA}

\date{August 1, 2007}

\begin{abstract}
  Strong repulsive interactions in a one-dimensional electron system
  suppress the exchange coupling $J$ of electron spins to a value much
  smaller than the Fermi energy $E_F$.  The conventional theoretical
  description of such systems based on the bosonization approach and the
  concept of Tomonaga-Luttinger liquid is applicable only at energies
  below $J$.  In this paper we develop a theoretical approach valid at all
  energies below the Fermi energy, including a broad range of energies
  between $J$ and $E_F$.  The method involves bosonization of the charge
  degrees of freedom, while the spin excitations are treated exactly.  We
  use this technique to calculate the spectral functions of strongly
  interacting electron systems at energies in the range $J\ll
  \varepsilon\ll E_F$.  We show that in addition to the expected features
  at the wavevector $k$ near the Fermi point $k_F$, the spectral function
  has a strong peak centered at $k=0$.  Our theory also provides
  analytical description of the spectral function singularities near
  $3k_F$ (the ``shadow band'' features).
\end{abstract}

\pacs{71.10.Pm 
%Fermions in reduced dimensions 
%(anyons, composite fermions, Luttinger liquid, etc.)
}

\maketitle

\section{Introduction}
\label{sec:introduction}

Recent experiments with quantum wires\cite{tarucha, yacoby, thomas,
  thomas1, cronenwett, kristensen, rokhinson, crook, thomas2, reilly1,
  yacoby1} and carbon nanotubes\cite{bockrath,yao} have stimulated
theoretical interest in transport properties of one-dimensional systems of
interacting electrons.  It is now widely accepted that in one dimension
interacting electrons form the so-called Luttinger
liquid.\cite{haldane,giamarchi} The main signature of the
Luttinger liquid---the power-law behavior of the tunneling density of
states---has recently been observed in experiments.\cite{yacoby1,
  bockrath, yao} Another well-known prediction\cite{maslov, ponomarenko,
  safi} of the Luttinger-liquid theory is that the conductance of a
quantum wire connecting two-dimensional leads should be quantized in units
of $2e^2/h$, regardless of the interaction strength.  Although the
quantization of conductance is routinely observed in modern experiments,
careful recent measurements show significant deviations\cite{thomas,
  thomas1, cronenwett, kristensen, rokhinson, crook, thomas2, reilly1}
from perfect quantization in the regime of very low electron density,
where the effective electron-electron interactions are very strong.

The applicability of the Luttinger-liquid theory is not expected to be
limited to weak interactions.  On the other hand, the properties of the
system do change significantly when the interactions become strong.  It is
well known that at low energies one-dimensional electron systems support
separate charge and spin excitation modes propagating at different
velocities,\cite{dzyaloshinskii} $v_\rho$ and $v_\sigma$.  Accordingly,
the Luttinger-liquid theory describes the low-energy excitations of the
system by two bosonic fields with linear dispersion, propagating at
velocities $v_\rho$ and $v_\sigma$.  The applicability of such a
description is limited to energies small compared to the bandwidths of the
charge and spin excitations $D_{\rho,\sigma}\sim\hbar v_{\rho,\sigma} n$,
where $n$ is the electron density.  In the non-interacting case, both
velocities coincide with the Fermi velocity $v_F$, so $D_\rho=D_\sigma\sim
E_F$.  In the presence of weak repulsive interactions the velocities are
renormalized, so that $v_\sigma<v_\rho$, but both velocities remain of
order $v_F$.  At strong interactions the spin mode velocity $v_\sigma$ is
strongly suppressed, $v_\sigma\ll v_\rho$.  In this case $D_\sigma\ll
E_F\lesssim D_\rho$, and the Luttinger-liquid theory is applicable only to
phenomena in which all the relevant energy scales are smaller than
$D_\sigma$.

A number of recent theory papers\cite{penc1, penc2, penc3, cheianov1,
  cheianov2, fiete1, fiete3, matveev1, matveev2, fiete2} addressed the
physics of strongly interacting electrons beyond the range of
applicability of the Luttinger-liquid theory.  Penc \textit{et
  al.}\cite{penc1, penc2, penc3} studied the tunneling density of states
and spectral functions of the one-dimensional Hubbard model at energies in
the range $D_\sigma\ll \varepsilon\ll D_\rho$ and zero temperature.
Cheianov and Zvonarev\cite{cheianov1,cheianov2} and Fiete and
Balents\cite{fiete1} explored the so-called spin incoherent
regime\cite{see_review} $D_\sigma\ll T\ll D_\rho$ and found an enhancement
of the tunneling density of states at energies $\varepsilon>T$.
Conductance of the quantum wire entering the spin-incoherent regime was
predicted\cite{matveev1, matveev2} to show behavior similar to the
anomalies observed in experiments.\cite{thomas, thomas1, cronenwett,
  kristensen, rokhinson, crook, thomas2, reilly1}

Despite the recent theoretical successes in treating strongly interacting
one-dimensional electrons, at present there is no regular theoretical
technique that can be applied to a broad class of problems and is not
limited to the exactly solvable models.\cite{penc1, penc2, penc3,
  cheianov1, cheianov2} In particular, the bosonization technique commonly
used to justify the Luttinger-liquid picture\cite{haldane,giamarchi} is
applicable only at energies below the spin bandwidth $D_\sigma$.  In this
paper we generalize the bosonization technique to all energies below the
\emph{charge} bandwidth $D_\rho$.  Our method treats the spin excitations
exactly, but applies bosonization to the charge excitations.  It is thus
applicable to a broad class of strong interactions, and is not limited to
short-range coupling required for the existence of the exact solutions.

We apply our technique to the calculation of the spectral functions and
the tunneling density of states of strongly interacting one-dimensional
electron systems.  Unlike many of the earlier treatments,\cite{penc1,
  penc2, penc3, cheianov1,cheianov2} our calculations can be applied to
systems with long-range interactions, such as quantum wires.  In addition,
our theory is valid in a broad range of temperatures: we obtain the
spectral functions at zero temperature, in the spin-incoherent case $T\gg
D_\sigma$, and also interpolate between these regimes.  Furthermore, our
approach provides a clear physical picture of the
enhancement\cite{penc1,cheianov1,fiete1} of the tunneling density of
states $\nu(\varepsilon)$ at $|\varepsilon|\ll E_F$.

Our approach is introduced in Sec.~\ref{sec:charge_bosonization}, where we
derive the expression for the electron field operators at energies below
$D_\rho$ by bosonizing the charge modes while treating the spin
excitations accurately, as excitations of an effective Heisenberg spin
chain with a small exchange constant $J$.  At energies below $D_\sigma\sim
J$ the spin excitations can also be bosonized.  This is accomplished in
Sec.~\ref{sec:spin_bosonization}, where we also demonstrate that the
standard bosonization expression for the electron
operators\cite{haldane,giamarchi} is recovered in our approach when all
important energy scales are below $J$.  Calculation of various physical
properties of the system requires knowledge of the electronic Green's
functions discussed in Sec.~\ref{sec:greens_functions}.  In the most
interesting regime of energies $|\varepsilon|\gg J$ the Green's functions
are expressed in terms of certain equal-time correlation functions
$c^\pm(q)$ of the Heisenberg spin chain.  Their behavior is important for
understanding the electronic transport at $|\varepsilon|\gg J$; it is
discussed in Sec.~\ref{sec:greens_functions}.  We calculate the spectral
functions at energies $J\ll|\varepsilon|\ll E_F$ in
Sec.~\ref{sec:spectral_functions} and show that their dependence on the
wavevector $k$ has a Gaussian peak centered at $k=0$, which determines the
behavior of the tunneling density of states.  In addition, we find
power-law singularities in the spectral functions at $k$ near $k_F$, as
well as the shadow band features\cite{penc2} near $3k_F$.  We conclude the
paper with the discussion of our results in Sec.~\ref{sec:discussion}.  A
brief summary of some of our results was reported in
Ref.~\onlinecite{brief}.

\section{Bosonization of charge excitations}
\label{sec:charge_bosonization}

The most experimentally relevant one-dimensional system of strongly
interacting electrons is realized in GaAs quantum wires.\cite{thomas,
  thomas1, cronenwett, kristensen, rokhinson, crook, thomas2, reilly1} In
these device the spectrum of electrons is quadratic, and the system can
be described by the standard Hamiltonian
\begin{eqnarray}
 \hspace{-1em}
  H&=&-\frac{\hbar^2}{2m_e}
     \int\! \psi_\gamma^\dagger(x)\partial_x^2\psi_\gamma(x)\, dx
\nonumber\\
   &&+\frac12\!\iint\!
      \psi_\gamma^\dagger(x)\psi_{\beta}^\dagger(y)
      V(x-y)
      \psi_{\beta}(y)\psi_\gamma(x) dxdy.
  \label{eq:original_Hamiltonian}     
\end{eqnarray}
Here $m_e$ is the effective mass of electrons, $\psi_\gamma(x)$ is the
annihilation operator of electron with spin $\gamma$ (summation over
repeating spin indices is implied), potential $V(x-y)$ describes the
interaction between electrons, and $\partial_x = d/dx$.

The assumption of quadratic spectrum in the Hamiltonian (1) is introduced
for simplicity, and most of the results we obtain apply to a generic
spectrum.  In particular, our theory is applicable to the one-dimensional
Hubbard model, with an important exception of the half-filled case, where
the charge excitation spectrum is gapped.

The effect of interactions on the low-energy properties of the system is
quantified by the parameters $\eta_f=V(0)/\hbar v_F$ and
$\eta_b=V(2k_F)/\hbar v_F$, where $V(q)$ is the Fourier transform of the
interaction potential.  Parameter $\eta_f$ controls the amplitude of
forward scattering of two electrons at the Fermi surface.  Positive value
of $\eta_f$ leads to the enhancement of the velocity of charge excitations
$v_\rho$ over the Fermi velocity.  Parameter $\eta_b$ controls the
amplitude of backward scattering of two electrons.  Strong backscattering
impedes propagation of spin excitations through the system and leads to
the suppression of the spin velocity $v_\sigma$.  In the case of
short-range interactions, the two parameters are of the same order of
magnitude.  On the other hand, in quantum wires the electrons interact via
long-range Coulomb repulsion, which is usually screened at a large
distance $d$ by a metal gate.  In this case $\eta_f/\eta_b\sim \ln(nd)$.
Throughout this paper we assume strong backscattering, $\eta_b\gg1$.

\subsection{Effective Hamiltonian}
\label{sec:effective_Hamiltonian}

In the limit $\eta_b\to \infty$ collisions of two electrons with opposite
spins result in complete backscattering.  In this case the processes of
spin exchange are completely suppressed, and the energy of the system no
longer depends on the spin degrees of freedom.  To find the energy of any
state, one can assume that all the spins $\gamma=\uparrow$, or,
equivalently, assume that the fermions in the Hamiltonian
(\ref{eq:original_Hamiltonian}) are spinless, $\psi_\gamma(x)\to\Psi(x)$.
The resulting Hamiltonian
\begin{eqnarray}
 \hspace{-1em}
  H_\rho&=&-\frac{\hbar^2}{2m_e}
     \int\! \Psi^\dagger(x)\partial_x^2\Psi(x)\, dx
\nonumber\\
   &&+\frac12\!\iint\!
      \Psi^\dagger(x)\Psi^\dagger(y)
      V(x-y)
      \Psi(y)\Psi(x)\, dxdy
  \label{eq:original_holon_Hamiltonian}
\end{eqnarray}
describes the charge excitations in the system.  Each eigenstate of the
Hamiltonian (\ref{eq:original_holon_Hamiltonian}) for a system of $N$
electrons is a degenerate multiplet of $2^N$ spin states.

In this paper the spinless fermions $\Psi(x)$ will be referred to as
\emph{holons.}  By construction their number equals the total number of
electrons,
\begin{equation}
  \label{eq:holon_density}
  \Psi^\dagger(x)\Psi(x)=\psi_\uparrow^\dagger(x)\psi_\uparrow^{}(x)
                        +\psi_\downarrow^\dagger(x)\psi_\downarrow^{}(x).
\end{equation}
In the limit of short-range coupling, $V(x-y)=V_0\delta(x-y)$, the
interactions in the Hamiltonian (\ref{eq:original_holon_Hamiltonian})
disappear due to the Pauli principle, $\Psi(x)\Psi(x)=0$, and holons
become free fermions.  This fact is well known in the theory of the
Hubbard model at strong interactions.\cite{ogata}

\begin{figure}[t]
 \resizebox{.46\textwidth}{!}{\includegraphics{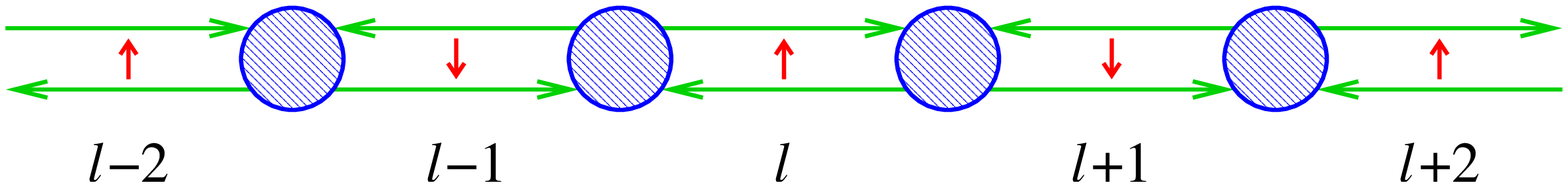}}
\caption{\label{fig:spin-regions} One-dimensional electrons at strong
  backscattering.  Each electron is confined to the region of space
  between the neighboring particles.  Shaded regions represent strong
  repulsive interactions.}
\end{figure}

The physical picture of one-dimensional electrons in the limit of strong
backscattering is illustrated in Fig.~\ref{fig:spin-regions}.  Due to the
strong repulsive interactions, electrons cannot pass through each other.
As a result electron $l$ is always confined between electrons $l-1$ and
$l+1$, and their spins cannot move through the system.

At strong but finite repulsion the amplitude of forward scattering of
electrons does not vanish.  Such processes give rise to a weak exchange of
the spins at the neighboring sites, e.g., $l$ and $l+1$.  To leading order
the coupling of the next-nearest neighbors can be neglected.  The symmetry
with respect to spin rotations dictates the form of coupling between the
spins:
\begin{equation}
  \label{eq:Heisenberg}
  H_\sigma = \sum_l J\, {\bm S}_{l}\cdot{\bm S}_{l+1}.
\end{equation}
The exchange constant $J$ is positive, as otherwise the ground state of a
system of one-dimensional electrons would have been spin-polarized, in
violation of the Lieb-Mattis theorem.\cite{lieb}

Thus we conclude that the Hamiltonian of a one-dimensional system of
strongly interacting electrons can be written as a sum $H_\rho +
H_\sigma$.  In the case of the Hubbard model this was first noticed by
Ogata and Shiba,\cite{ogata} who discovered that in the limit of strong
repulsion the Bethe ansatz ground state of the system factorizes into a
direct product of the ground state of non-interacting fermions (holons)
and the ground state of the Heisenberg spin chain (\ref{eq:Heisenberg}).
In the context of quantum wires the description based on the Hamiltonian
$H_\rho + H_\sigma$ was used in Refs.~\onlinecite{matveev1,matveev2}.  The
long-range nature of the Coulomb repulsion between electrons in a quantum
wire results in the exponential suppression of the exchange
constant,\cite{hausler,matveev2,klironomos,fogler}
\begin{equation}
  \label{eq:J}
  J=J^*\exp\left(-\frac{2.80}{\sqrt{na_B}}\right).
\end{equation}
Here the prefactor $J^*\sim E_F(na_B)^{-3/4}$, the Bohr radius is defined
as $a_B=\epsilon\hbar^2/m_ee^2$, and $\epsilon$ is the dielectric constant.

It is worth mentioning that one can replace the Hamiltonian
(\ref{eq:original_Hamiltonian}) with the sum of two independent
Hamiltonians (\ref{eq:original_holon_Hamiltonian}) and
(\ref{eq:Heisenberg}) only at sufficiently low energies.  Indeed, the
backscattering amplitude for two electrons with wavevectors $\pm k$ is
$V(2k)/\hbar v_k$, and tends to zero at $k\to \infty$ for any reasonable
interaction potential. (Here $v_k=\hbar k/m$ is the velocity of an
electron with wavevector $k$.)  Thus at high energies the initial
assumption of strong backscattering is violated.  To find the region of
applicability of our low-energy theory, one can estimate the correction to
the exchange constant (\ref{eq:J}) caused by the fact that at higher
energies the distances between electrons fluctuate, and the density $n$ is
no longer constant.  Given that the rigidity of the Wigner crystal is due
to the Coulomb repulsion between the electrons, we estimate $\delta n \sim
\sqrt{m_ena_B|\varepsilon|}/\hbar$, and the exchange $J$ acquires
significant corrections in the presence of excitations with energies
$|\varepsilon|\gtrsim (\hbar n)^2/m_e\sim E_F$.  Thus our subsequent
results are valid up to energies of order $E_F$, rather than the somewhat
higher energy scale $D_\rho\sim E_F\sqrt{\ln(nd)/na_B}$.  In the case of
short-range interactions the scales $D_\rho$ and $E_F$ are of the same
order of magnitude.

\subsection{Electron creation and destruction operators}
\label{sec:electron_operators}

The effective Hamiltonian (\ref{eq:original_holon_Hamiltonian}),
(\ref{eq:Heisenberg}) is defined in terms of the holon field operators
$\Psi(x)$ and the spin operators ${\bm S}_l$, rather than the original
electron operators $\psi_{\uparrow,\downarrow}(x)$.  In order to apply the
effective theory (\ref{eq:original_holon_Hamiltonian}),
(\ref{eq:Heisenberg}) to problems formulated in terms of electrons (e.g.,
calculation of the spectral functions) we need to establish the relations
between the electron operators and the new variables.

In the context of the Hubbard model this issue was addressed by Penc
\emph{et al.},\cite{penc1} who used the definition
\begin{equation}
  \label{eq:Penc_operator}
  \psi_\gamma^\dagger(0)=\Psi^\dagger(0) Z^\dagger_{0,\gamma}.
\end{equation}
Given the relation (\ref{eq:holon_density}) between the densities of
electrons and holons, creation of an electron at point 0 must be
accompanied by creation of a holon.  In addition, when a new particle is
added to the system of $N$ electrons, the spin chain (\ref{eq:Heisenberg})
acquires an additional site.  This is accounted for by the operator
$Z^\dagger_{0,\gamma}$.  By definition, operator $Z_{l, \gamma}^\dagger$
adds a new site with spin $\gamma$ to the spin chain between the sites
$l-1$ and $l$.

Despite the fact that the rule (\ref{eq:Penc_operator}) leads to a number
of correct results when applied carefully,\cite{penc1, penc2, penc3} it is
not a completely satisfactory expression of an electron creation operator
in terms of the charge and spin degrees of freedom.  In particular, the
generalization of Eq.~(\ref{eq:Penc_operator}) to $x\neq0$ is not
straightforward.\cite{penc2} The origin of the difficulty lies in the fact
that unlike the electrons and holons, the spins in the Hamiltonian
(\ref{eq:Heisenberg}) are not assigned to specific points in space.  When
an electron is created at point $x$, the additional site in the spin chain
(\ref{eq:Heisenberg}) appears at $l=l(x)$, where 
\begin{equation}
  l(x)=\int_{-\infty}^{x+0} \!\Psi^\dagger(y)\Psi(y)\, dy
\label{eq:l(x)}
\end{equation}
is the number of electrons (or holons) between $-\infty$ and point $x$. We
shall therefore define the electron creation and annihilation operators as
\begin{subequations}
  \label{eq:our_operators}
\begin{eqnarray}
  \label{eq:our_creation_operator}
  \psi_\gamma^\dagger(x)&=&Z^\dagger_{l(x),\gamma}\Psi^\dagger(x),
\\
  \label{eq:our_annihilation_operator}
  \psi_\gamma(x)&=&\Psi(x)Z_{l(x),\gamma}.
\end{eqnarray}
\end{subequations}
The most important difference between the expressions
(\ref{eq:Penc_operator}) and (\ref{eq:our_creation_operator}) is that the
latter explicitly accounts for the fact that the spins are attached to
electrons.  Thus despite the apparent separation of the charge and spin
degrees of freedom in the effective Hamiltonian
(\ref{eq:original_holon_Hamiltonian}), (\ref{eq:Heisenberg}), the electron
creation operator (\ref{eq:our_creation_operator}) does not factorize into
a product of two operators acting on only charge or only spin
variables.\cite{footnote1} In Appendix~\ref{sec:anticommutation} we show
that our operators (\ref{eq:our_operators}) satisfy the appropriate
anticommutation relations.

\subsection{Bosonization of holon operators}
\label{sec:holon_bosonization}

In this paper we are interested in the properties of strongly-interacting
one-dimensional systems at energies well below the Fermi energy,
$|\varepsilon|\ll E_F$.  In this case the dynamics of the charge degrees
of freedom described by the Hamiltonian
(\ref{eq:original_holon_Hamiltonian}) simplifies dramatically.  Indeed, it
is well known that the low-energy properties of a system of interacting
spinless fermions are accurately described by the Tomonaga-Luttinger
model,\cite{haldane,giamarchi}
\begin{equation}
  \label{eq:bosonized_holon_Hamiltonian}
  H_\rho=\frac{\hbar v_\rho}{2\pi}
         \int\left[K(\partial_x\theta)^2+K^{-1}(\partial_x\phi)^2\right]dx.
\end{equation}
Here $\phi$ and $\theta$ are bosonic fields satisfying the commutation
relations $[\phi(x),\partial_y\theta(y)]=\pi i\,\delta(x-y)$.  Field $\theta$
is related to the momentum density of the system, $p(x)=\hbar
n\,\partial_x\theta(x)$, whereas $\phi$ is defined in terms of the density
of fermions,
\begin{equation}
  \label{eq:bosonized_density}
  \Psi^\dagger(x)\Psi(x) = \frac{1}{\pi}\big[k_F^h + \partial_x\phi(x)\big].
\end{equation}
Here we have introduced the Fermi momentum of the holons $k_F^h$.
Considering that the total electron density equals the holon density,
Eq.~(\ref{eq:holon_density}), we have $n=2k_F/\pi=k_F^h/\pi$.  Thus the
Fermi wavevectors of holons and electrons are related by
\begin{equation}
  \label{eq:holon_Fermi_wavevector}
  k_F^h=2k_F.
\end{equation}
Finally, the parameter $K\leq1$ in the Hamiltonian
(\ref{eq:bosonized_holon_Hamiltonian}) is the so-called Luttinger-liquid
constant.

Transformation of the Hamiltonian $H_\rho$ from the fermionic form
(\ref{eq:original_holon_Hamiltonian}) to the bosonic form
(\ref{eq:bosonized_holon_Hamiltonian}) is accomplished via bosonization
procedure,\cite{haldane,giamarchi} in which the fermion operators $\Psi$
are expressed in terms of the bosonic fields $\phi$ and $\theta$.  At the
first step one notices that at low energies the properties of
one-dimensional Fermi systems are dominated by excitations near the two
Fermi points.  Particles near each of the Fermi points propagate in one of
two possible directions, right or left.  Thus the fermion operator is
presented as a sum of two chiral fermions,
\begin{equation}
  \label{eq:left-right}
  \Psi(x)=\Psi_R(x)+\Psi_L(x),
\end{equation}
where the operators $\Psi_R$ and $\Psi_L$ destroy fermions with
wavevectors near the right and left Fermi points, respectively.

The chiral fermion operators are bosonized following the prescription
\begin{equation}
  \label{eq:holon_bosonization}
  \Psi_{R,L}(x)=\frac{1}{\sqrt{2\pi\alpha}}\, e^{-i\theta(x)}
                e^{\pm i[k_F^h x+\phi(x)]},
\end{equation}
where $\alpha$ is a short-distance cut-off.  

Using the relations (\ref{eq:l(x)}), (\ref{eq:bosonized_density}),
(\ref{eq:left-right}), and (\ref{eq:holon_bosonization}), we express the
electron annihilation operator (\ref{eq:our_annihilation_operator}) in the
form
\begin{eqnarray}
  \psi_{\gamma}(x)&=&\frac{e^{-i\theta(x)}}{\sqrt{2\pi\alpha}}\, 
                     \left(e^{i[k_F^h x+\phi(x)]} 
                           + e^{- i[k_F^h x+\phi(x)]}\right)
\nonumber\\
                  &&\times Z_{l,\gamma}\Big|_{l=\frac{1}{\pi}[k_F^hx+\phi(x)]}.
  \label{eq:our_annihilation_operator_bosonized}
\end{eqnarray}
Unlike the original formula (\ref{eq:our_annihilation_operator}), this
expression is valid only at low energies, $|\varepsilon|\ll E_F$.  Its
advantage is that the charge modes are now presented in the form of
non-interacting bosons (\ref{eq:bosonized_holon_Hamiltonian}) and can be
treated rather easily.  We apply the expression
(\ref{eq:our_annihilation_operator_bosonized}) to the calculation of the
electron Green's functions at $J,T\ll |\varepsilon|\ll E_F$ in
Sec.~\ref{sec:greens_functions}.

\section{Bosonization of spin excitations}
\label{sec:spin_bosonization}

As we discussed in Sec.~\ref{sec:introduction}, the conventional
Tomonaga-Luttinger theory\cite{haldane,giamarchi} of low-energy properties
of one-dimensional electron systems is based on the idea of bosonization
of electron operators.  Mathematically this corresponds to applying the
procedure (\ref{eq:holon_bosonization}) to annihilation operators of
electrons with spins $\gamma=\uparrow,\downarrow$,
\begin{subequations}
  \label{eq:electron_bosonization}
\begin{eqnarray}
    \psi_{R\gamma}(x)
      &=&\frac{e^{ik_Fx}}{\sqrt{2\pi\alpha}} 
                 e^{\frac{i}{\sqrt2}[\phi_\rho(x)-\theta_\rho(x)]}
\nonumber\\
      &&\times   e^{\pm\frac{i}{\sqrt2}[\phi_\sigma(x)-\theta_\sigma(x)]},
  \label{eq:electron_bosonization_right}
\\[1ex]
  \psi_{L\gamma}(x)
      &=&\frac{e^{-ik_Fx}}{\sqrt{2\pi\alpha}} 
                 e^{-\frac{i}{\sqrt2}[\phi_\rho(x)+\theta_\rho(x)]}
\nonumber\\
      &&\times
  e^{\mp\frac{i}{\sqrt2}[\phi_\sigma(x)+\theta_\sigma(x)]}.
  \label{eq:electron_bosonization_left}
\end{eqnarray}
\end{subequations}
Here $\phi_{\rho,\sigma}$ and $\theta_{\rho,\sigma}$ are the bosonic
fields describing the charge and spin excitation modes of the system.  In
terms of these fields the Hamiltonian of interacting electrons
(\ref{eq:original_Hamiltonian}) takes its bosonized low-energy form
$H_\rho+H_\sigma$ with
\begin{subequations}
\label{eq:standard_bosonization}
\begin{eqnarray}
\label{eq:H_rho}
  H_\rho&=&\frac{\hbar v_\rho}{2\pi}\int
                \left[K_\rho (\partial_x\theta_\rho)^2 
                 +K_\rho^{-1}(\partial_x\phi_\rho)^2\right]dx,
\\
  H_\sigma&=&\frac{\hbar v_\sigma}{2\pi}\int
                \left[K_\sigma (\partial_x\theta_\sigma)^2 
                 +K_\sigma^{-1}(\partial_x\phi_\sigma)^2\right]dx
\nonumber\\
          &&+\frac{2g_{1\perp}}{(2\pi\alpha)^2}\int 
             \cos\left[\sqrt8\, \phi_\sigma(x)\right] dx.
\label{eq:H_sigma}
\end{eqnarray}
\end{subequations}
\noindent
Here the matrix element $g_{1\perp}$ accounts for the processes of
backscattering of two electrons with opposite spins.  The respective
sine-Gordon term in the Hamiltonian (\ref{eq:H_sigma}) is marginally
irrelevant.  In the absence of magnetic field, the SU(2) symmetry of the
problem requires\cite{giamarchi} that when $g_{1\perp}$ scales to zero,
the Luttinger parameter $K_\sigma\to1$.

The Hamiltonian (\ref{eq:standard_bosonization}) is typically derived
under the assumption that the electron-electron interactions are weak.  On
the other hand, it represents a stable low-energy fixed point of the
theory, and thus should be valid beyond the weak-interaction
approximation.  Although the nature of the low-energy fixed point can, in
principle, change at a finite value of the interaction strength, such a
change would imply a quantum phase transition, which is generally not
expected.  The more likely scenario is that Hamiltonian
(\ref{eq:standard_bosonization}) is the correct low-energy description of
one-dimensional electron systems at arbitrarily strong interactions.
Under this assumption one should expect to be able to show that at low
energies (i) the Hamiltonian (\ref{eq:bosonized_holon_Hamiltonian}),
(\ref{eq:Heisenberg}) is equivalent to
Eq.~(\ref{eq:standard_bosonization}), and (ii) at the same time our
expression (\ref{eq:our_annihilation_operator_bosonized}) for the electron
destruction operator transforms to Eq.~(\ref{eq:electron_bosonization}).
We now show that this is indeed the case.

\subsection{Low-energy Hamiltonian of strongly-interacting electrons in
  one dimension}
\label{sec:relation_between_Hamiltonians}

First we notice that the expressions
(\ref{eq:bosonized_holon_Hamiltonian}) and (\ref{eq:H_rho}) for $H_\rho$
are very similar, as they both describe acoustic excitations in the charge
channel propagating at speed $v_\rho$.  Although it is natural to assume
that $K$, $\phi$, and $\theta$ in
Eq.~(\ref{eq:bosonized_holon_Hamiltonian}) should be identified with
$K_\rho$, $\phi_\rho$, and $\theta_\rho$ in Eq.~(\ref{eq:H_rho}),
respectively, this is not the case.  The correct approach is to ensure
that the physically observable quantities, such as the electron density
(\ref{eq:holon_density}), have equivalent expressions in both theories.
In the standard bosonization description based on
Eq.~(\ref{eq:electron_bosonization}) the electron density is given by
\begin{equation}
  \label{eq:electron_density}
  \psi_\uparrow^\dagger(x)\psi_\uparrow^{}(x)
   +\psi_\downarrow^\dagger(x)\psi_\downarrow^{}(x)
  = \frac{2k_F}{\pi} + \frac{\sqrt2}{\pi}\,\partial_x\phi_\rho(x).
\end{equation}
Comparing this expression with Eqs.~(\ref{eq:holon_density}),
(\ref{eq:bosonized_density}), and (\ref{eq:holon_Fermi_wavevector}), we
conclude
\begin{subequations}
  \label{eq:relations_between_charge_variables}
  \begin{equation}
    \label{eq:phi-phi_rho}
    \phi(x)=\sqrt{2}\, \phi_\rho(x).
  \end{equation}
  Then, to preserve the proper commutation relations between bosonic
  fields, one has to assume
  \begin{equation}
    \label{eq:theta-theta_rho}
    \theta(x)=\frac{1}{\sqrt{2}}\, \theta_\rho(x).
  \end{equation}
Finally, substituting Eqs.~(\ref{eq:phi-phi_rho}) and
(\ref{eq:theta-theta_rho}) into the Hamiltonian
(\ref{eq:bosonized_holon_Hamiltonian}), we recover Eq.~(\ref{eq:H_rho})
if
\begin{equation}
  \label{eq:K-K_rho}
  K=2K_\rho.
\end{equation}
\end{subequations}

Turning to the spin Hamiltonian $H_\sigma$, we note that there is a
well-known procedure\cite{giamarchi} of bosonization of the Heisenberg
model (\ref{eq:Heisenberg}).  One starts by converting spin operators
${\bm S}_l$ to spinless fermion operators $a_l$ via the Jordan-Wigner
transformation
\begin{equation}
  \label{eq:Jordan-Wigner}
  S_l^z = a_l^\dagger a_l - \frac12,
  \quad
  S_l^x + iS_l^y 
      = a_l^\dagger\exp\left(i\pi\sum_{j=1}^{l-1} a_j^\dagger a_j\right).
\end{equation}
In terms of the spinless fermions the Hamiltonian $H_\sigma$ takes the
form
\begin{subequations}
  \label{eq:Heisenberg_fermionized}
\begin{eqnarray}
    H_\sigma &=& H^{xy}+H^z,
  \label{eq:xy+z}
\\
    H^{xy}
     &=&\frac12 \sum_l J
        \left(a_l^\dagger a_{l+1}^{} + a_{l+1}^\dagger a_{l}^{}\right),
  \label{eq:H^xy}
\\
    H^z
     &=&\sum_l J \left(a_l^\dagger a_l^{} - \frac12 \right)
                 \left(a_{l+1}^\dagger a_{l+1}^{} - \frac12 \right).
  \label{eq:H^z}
\end{eqnarray}
\end{subequations}
Thus the Heisenberg model (\ref{eq:Heisenberg}) is reduced to the
tight-binding model of spinless fermions with repulsive interactions
between particles at the nearest-neighbor sites.

The steps leading from Eq.~(\ref{eq:Heisenberg}) to
(\ref{eq:Heisenberg_fermionized}) are exact, and the spectra of the two
Hamiltonians are identical at all energies.  At energies much smaller than
$J$ one can simplify the Hamiltonian (\ref{eq:Heisenberg_fermionized}) by
bosonizing the Jordan-Wigner fermions.  One starts by considering the
non-interacting model given by Eq.~(\ref{eq:H^xy}).  The spectrum of that
Hamiltonian is obtained as the sum of energies $\epsilon(q)$ of
independent spinless fermions,
\begin{equation}
  \label{eq:tight_binding_spectrum}
  \epsilon(q)= J\cos ql,
\end{equation}
where the wavevector $q$ varies from $0$ to $2\pi$.  

In the absence of external magnetic field, one expects $\langle
S_l^z\rangle=0$.  According to Eq.~(\ref{eq:Jordan-Wigner}) the band
(\ref{eq:tight_binding_spectrum}) is half-filled, and the two Fermi points
are at
\begin{equation}
  \label{eq:Fermi_points}
  q_L=\frac{\pi}{2},
\quad
  q_R=\frac{3\pi}{2}.
\end{equation}
The bosonization is accomplished by presenting the operator $a_l$ as a sum
of operators destroying the right- and left-moving particles,
\begin{equation}
  \label{eq:right-left}
  a_l = a_R(l) + a_L(l),
\end{equation}
where
\begin{subequations}
\label{eq:chiral_fermions}
  \begin{eqnarray}
  \label{eq:psi_R}
  a_R(\xi) 
     &=& \frac{1}{\sqrt{2\pi\tilde\alpha}}\, 
         e^{iq_R\xi} e^{i\varphi_R(\xi)},
  \\
  \label{eq:psi_L}
  a_L(\xi) 
     &=& \frac{1}{\sqrt{2\pi\tilde\alpha}}\, 
         e^{iq_L\xi} e^{-i\varphi_L(\xi)}.
\end{eqnarray}
\end{subequations}
Since the bosonization description concentrates on the range of momenta
close to the Fermi points, the discrete site number $l$ is replaced here
with the continuous coordinate $\xi$.  (Unlike the coordinate $x$ of
electrons, $\xi$ is dimensionless.)  The chiral bosonic fields
$\varphi_R$ and $\varphi_L$ satisfy the commutation relations
\begin{subequations}
\label{eq:chiral_boson_commutation}
\begin{eqnarray}
  [\varphi_R(\xi), \partial_{\xi'}\varphi_R(\xi')]
     &=&-2\pi i\,\delta(\xi-\xi'),
\\{}
  [\varphi_L(\xi), \partial_{\xi'}\varphi_L(\xi')]
     &=&2\pi i\,\delta(\xi-\xi'),
\\{}
  [\varphi_L(\xi),\partial_{\xi'}\varphi_R(\xi')]&=&0.
\end{eqnarray}
\end{subequations}

Upon the bosonization (\ref{eq:chiral_fermions}) the Hamiltonian
(\ref{eq:H^xy}) takes the form
\begin{equation}
  \label{eq:H^xy_bosonized}
  H^{xy} =
  \frac{J}{4\pi}
  \int[(\partial_{\xi}\varphi_R)^2+(\partial_{\xi}\varphi_L)^2]d\xi.
\end{equation}
The next step of the bosonization procedure is to convert to nonchiral
bosonic fields
\begin{equation}
  \label{eq:non-chiral_bosons}
    \varphi = \frac12 (\varphi_L + \varphi_R),
  \quad
  \vartheta = \frac12 (\varphi_L - \varphi_R).
\end{equation}
As a result the Hamiltonian (\ref{eq:H^xy_bosonized}) takes the form
\begin{subequations}
  \label{eq:Heisenberg_bosonized}
\begin{equation}
    \label{eq:H_0}
    H_0 = \frac{\hbar \tilde v}{2\pi}\int
          \left[\mathcal{K} (\partial_{\xi}\vartheta)^2 
          +\frac1{\mathcal K}(\partial_{\xi}
                                   \varphi)^2\right]d\xi,
\end{equation}
with $\tilde v=J/\hbar$ and $\mathcal{K}=1$.

When the interaction term (\ref{eq:H^z}) is added to the Hamiltonian
(\ref{eq:H^xy}), and the bosonization transformation
(\ref{eq:chiral_fermions}) is applied, the parameters $\tilde v$ and
$\mathcal{K}$ change, and an additional term appears in the Hamiltonian,
\begin{equation}
  \label{eq:V}
  V=\frac{2g}{(2\pi\tilde\alpha)^2}\int 
             \cos\left[4 \varphi(\xi)\right] d\xi,
\end{equation}
\end{subequations}
with $g\sim1$.  

The Hamiltonian (\ref{eq:Heisenberg_bosonized}) is equivalent to the spin
part (\ref{eq:H_sigma}) of the Hamiltonian of weakly-interacting electron
system, if one assumes
\begin{subequations}
  \label{eq:relations_between_spin_variables}
\begin{eqnarray}
  \varphi(\xi) &=& \frac{1}{\sqrt2}\,\phi_\sigma(\xi/n),
\\
  \vartheta(\xi) &=& \sqrt2\,\theta_\sigma(\xi/n),
\\
\tilde v &=& v_\sigma n,
\\
\mathcal{K} &=& \frac{K_\sigma}{2}.
\label{eq:K_sigma_relation}
\end{eqnarray}
\end{subequations}
In the absence of magnetic field, as the cosine term (\ref{eq:V}) scales
to zero at low energies, the Luttinger parameter $\mathcal{K}$ approaches
$1/2$, as required by the SU(2) symmetry of the problem.\cite{giamarchi}
Thus Eq.~(\ref{eq:K_sigma_relation}) is consistent with the similar
requirement $K_\sigma\to1$ in the Hamiltonian (\ref{eq:H_sigma}).

\subsection{Bosonization of the operators $Z_{l,\gamma}$}
\label{sec:Z-bosonization}

In order to demonstrate that in the regime of low energies
$|\varepsilon|\ll J$ the electron destruction operator
(\ref{eq:our_annihilation_operator_bosonized}) takes the standard form
(\ref{eq:electron_bosonization}), one needs to bosonize the operator
$Z_{l,\gamma}$ in Eq.~(\ref{eq:our_annihilation_operator_bosonized}).  We
start with $Z_{0,\downarrow}$.  This operator acts on an arbitrary state
in the Hilbert space of the Hamiltonian (\ref{eq:Heisenberg}) with $N$
sites.  If the spin at the site $l=0$ is $\downarrow$, the operator
$Z_{0,\downarrow}$ removes that site from the spin chain; if the spin is
$\uparrow$, the outcome is zero.  The Jordan-Wigner transformation
(\ref{eq:Jordan-Wigner}) defines one spinless fermion per each site with
spin $\uparrow$.  Thus the operator $Z_{0,\downarrow}$ removes site $l=0$
from the tight-binding model (\ref{eq:H^xy}) without changing the number
of fermions $N_f$.

To derive the bosonized form of $Z_{0,\downarrow}$ it is convenient to
consider its effect upon the eigenstates of the non-interacting model
(\ref{eq:H^xy}).  The latter are Slater determinants of plane waves with
wavevectors in the range $0\leq q_j<2\pi$ and energies $\epsilon(q_j)$
given by Eq.~(\ref{eq:tight_binding_spectrum}).  To determine the allowed
values of wavevectors $q_j$ we assume periodic boundary conditions on the
spin chain, ${\bm S}_0={\bm S}_N$.  Because of the Jordan-Wigner string in
the definition (\ref{eq:Jordan-Wigner}) of the spinless fermions, their
respective boundary conditions are either periodic, or antiperiodic,
depending on the parity of their number $N_f$,
\begin{equation}
  \label{eq:fermion_boundary_conditions}
  a_N^\dagger=(-1)^{N_f-1}a_0^\dagger.
\end{equation}
Thus the wavevectors of the fermions take the values
\begin{equation}
  \label{eq:fermion_wavevectors}
  q_j=\left\{
      \begin{array}[c]{ll}
       \frac{2\pi}{N}j, &  \mbox{for odd $N_f$,}\\[1ex]
       \frac{2\pi}{N}\big(j+\frac12\big),\ &  \mbox{for even $N_f$,}
      \end{array}
      \right.
\end{equation}
where $j= 0, 1,\ldots,N-1$.  

\begin{figure}[t]
 \resizebox{.4\textwidth}{!}{\includegraphics{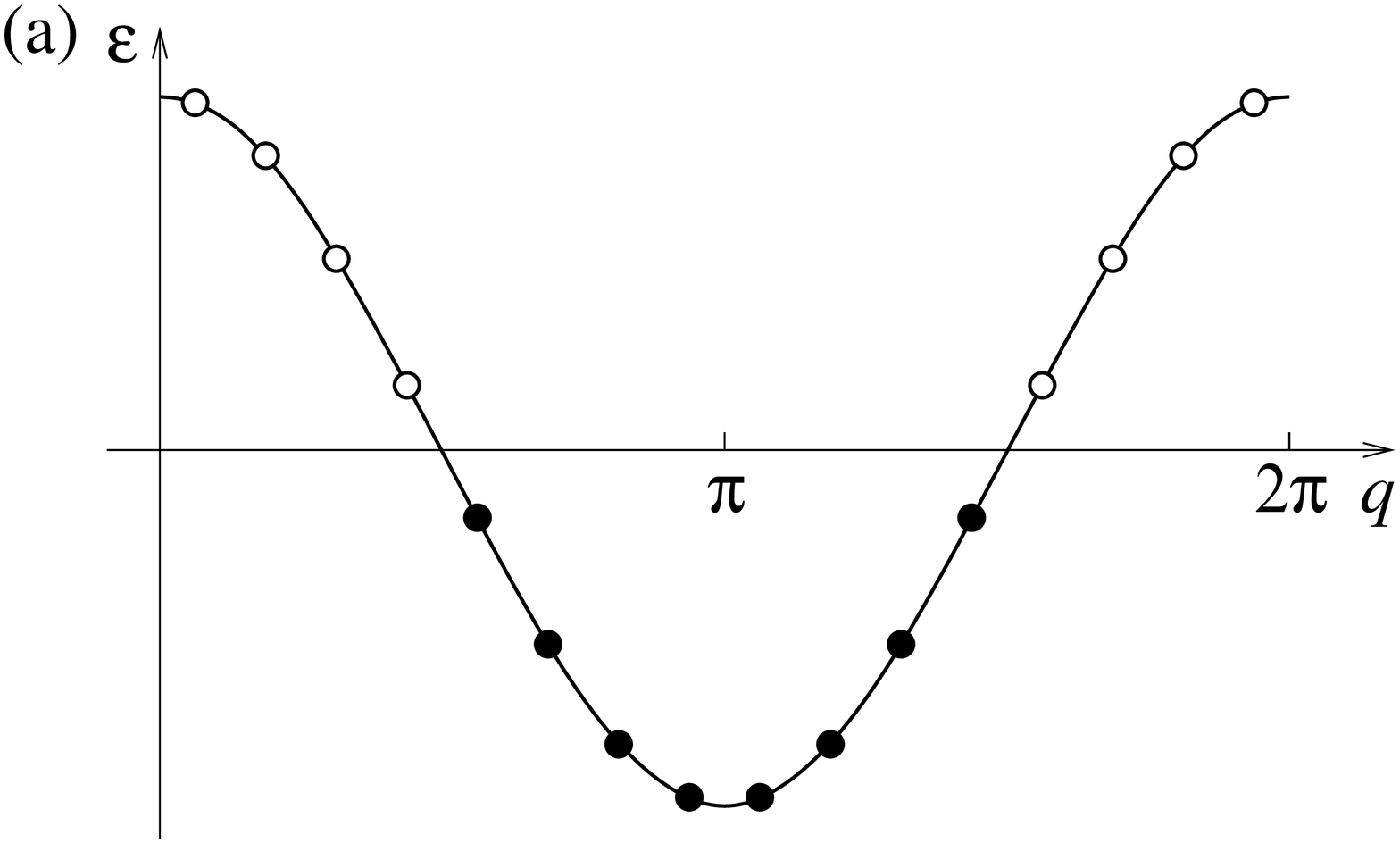}}
\\[2ex]
 \resizebox{.4\textwidth}{!}{\includegraphics{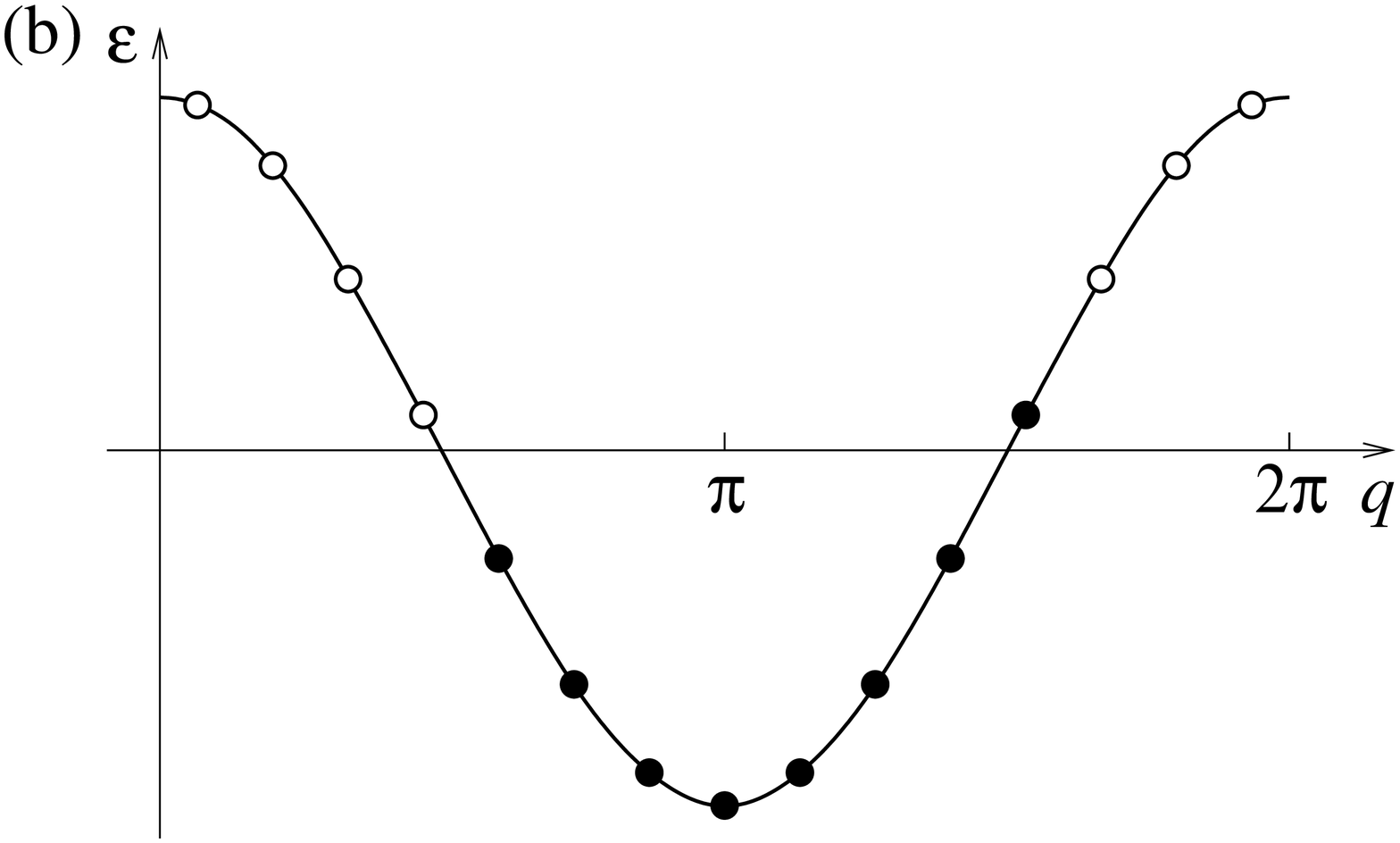}}
\caption{\label{fig:groundstate} Ground state of the system of
  noninteracting fermions (\ref{eq:H^xy}) for the case of 16 sites and 8
  fermions.  Filled circles show wavevectors and energies of the 8
  occupied single-particle states.  (b) Ground state of the same problem
  with 15 sites.}
\end{figure}

The ground state of the Hamiltonian (\ref{eq:H^xy}) is illustrated in
Fig.~\ref{fig:groundstate}(a).  Upon removal of one site from the chain
the allowed values of the wavevector increase slightly,
\begin{equation}
  \label{eq:momentum_change}
  q_j\to q_j\frac{N}{N-1}\simeq q_j+\frac{q_j}{N}.
\end{equation}
The effect of the operator $Z_{0,\downarrow}$ on the eigenstates of
$H^{xy}$ can be interpreted as follows.  By removing a site from the spin
chain, $Z_{0,\downarrow}$ creates a scattering potential for the
Jordan-Wigner fermions near $\xi=0$.  Only forward scattering is
present, and the wavefunctions of the fermions acquire phase shifts
proportional to the wavevectors,
\begin{equation}
  \label{eq:wavefunction_transformation}
  e^{iq\xi}\to 
         e^{iq\xi} e^{-i\frac{q}{2}{\rm sgn}\,(\xi)}.
\end{equation}
When the periodic boundary conditions are imposed on the fermions, the
phase shift in Eq.~(\ref{eq:wavefunction_transformation}) moves the
allowed values of the wavevectors by $q/N$, in agreement with
Eq.~(\ref{eq:momentum_change}).

In the bosonization treatment of the spin chain one concentrates on the
vicinities of the two Fermi points $q_R$ and $q_L$, where the fermions are
classified as either right- or left-moving, Eq.~(\ref{eq:right-left}).
According to Eq.~(\ref{eq:wavefunction_transformation}), the operator
$Z_{0,\downarrow}$ transforms the fermion operators as
\begin{equation}
  \label{eq:operator_transformation}
  a_{R,L}(\xi) \to 
  Z_{0,\downarrow} a_{R,L}(\xi) Z_{0,\downarrow}^\dagger
  =a_{R,L}(\xi) e^{-i\frac{q_{R,L}^{}}{2}\,{\rm sgn\,}(\xi)}.
\end{equation}
Using the bosonized representation (\ref{eq:chiral_fermions}) of the
fermion operators and the commutation relations
(\ref{eq:chiral_boson_commutation}), we conclude 
\begin{equation}
  \label{eq:Z_0down}
  Z_{0,\downarrow}=
   \exp\left\{-\frac{i}{2\pi}
             \big[q_L^{}\varphi_L^{}(0)+q_R^{}\varphi_R^{}(0)\big]\right\}.
\end{equation}
Here we omit a numerical prefactor, which depends on the specific cut-off
procedure used in the bosonization scheme, but can be considered to be of
order unity.

To find the bosonized expression for $Z_{l,\downarrow}$ away from the
point $l=0$ it is not sufficient to replace the arguments of $\varphi_L$
and $\varphi_R$ with $l$.  Indeed, our derivation of
Eq.~(\ref{eq:Z_0down}) allowed for an arbitrary phase factor, which may
depend on $l$.  To determine this phase factor, we notice that
\begin{equation}
  \label{eq:Z_l}
  Z_{l,\gamma} = e^{-i\hat Q l} Z_{0,\gamma} e^{i\hat Q l},
\end{equation}
where $\hat Q$ is the operator of the total momentum of the system.  (This
relation becomes clear if one notices that the operator $e^{i\hat Q l}$
shifts the spin chain by $l$ sites to the left.)  Summing the changes of
wavevectors (\ref{eq:momentum_change}) for all particles between the Fermi
points (\ref{eq:Fermi_points}), we conclude $Z_{0,\sigma}$ increases $Q$
by $\pi/2$.  Thus we obtain
\begin{equation}
  \label{eq:Z_ldown_v1}
  Z_{l,\downarrow}=\exp\left[-i\,\frac{\pi}{2}\,l
                   -\frac{i}{4}\,\varphi_L^{}(l)
                   -\frac{3i}{4}\,\varphi_R^{}(l)\right],
\end{equation}
where we have also substituted into Eq.~(\ref{eq:Z_0down}) the values of
the Fermi wavevectors (\ref{eq:Fermi_points}).

The apparent asymmetry between the left- and right-movers in
Eq.~(\ref{eq:Z_ldown_v1}) can be understood by noticing that the
interpretation (\ref{eq:momentum_change}) of the changes of wavevectors of
the fermions is not unique.  Instead of assuming that as we remove a site
from the spin chain, the fermion states in Fig.~\ref{fig:groundstate}(a)
transform into those of Fig.~\ref{fig:groundstate}(b) by shifting to the
right, Eq.~(\ref{eq:momentum_change}), one can assume that all the states
move to the left,
\begin{equation}
  \label{eq:momentum_change_alternative}
  q_j\to q_j\frac{N}{N-1}-\frac{2\pi}{N-1}\simeq q_j+\frac{q_j-2\pi}{N}.
\end{equation}
As a result, the system arrives at a new ground state on the $(N-1)$-site
lattice, which is the mirror image of the state shown in
Fig.~\ref{fig:groundstate}(b).  One can repeat the above arguments leading
to Eq.~(\ref{eq:Z_0down}) and obtain the new expression by replacing
$q_{R,L}\to q_{R,L}-2\pi$.  Similar to the presentation of the fermion
operators $a_l$ as a sum of two chiral contributions
(\ref{eq:right-left}), we conclude
\begin{eqnarray}
  Z_{l,\downarrow}&=&\exp\left[-i\,\frac{\pi}{2}\,l
                      -\frac{i}{4}\,\varphi_L^{}(l)
                      -\frac{3i}{4}\,\varphi_R^{}(l)\right]
\nonumber\\
                  &&+\exp\left[i\,\frac{\pi}{2}\,l
                      +\frac{3i}{4}\,\varphi_L^{}(l)
                      +\frac{i}{4}\,\varphi_R^{}(l)\right].
  \label{eq:Z_ldown}
\end{eqnarray}

Apart from these two contributions, the operator $Z_{l,\downarrow}$ may
contain terms corresponding to greater shifts of the fermion states in the
momentum space.  They can be obtained by adding any multiples of $2\pi/N$
to the right-hand side of Eq.~(\ref{eq:momentum_change}).  Such terms can
be viewed as operator (\ref{eq:Z_0down}) combined with
$(\mbox{$a_L$\!}^\dagger a_R)^m$ or $(\mbox{$a_R$\!}^\dagger a_L)^m$.  At
low energies such terms are less relevant than the leading contributions
(\ref{eq:Z_ldown}), and can be neglected.

A similar bosonization procedure can be performed with operator
$Z_{l,\uparrow}$.  In addition to the shifts (\ref{eq:momentum_change}) of
the wavevectors of the allowed fermion states caused by the change of
system size $N\to N-1$, one also needs to account for the change in the
fermion number, $N_f\to N_f-1$.  The latter changes the wavevector
quantization conditions (\ref{eq:fermion_wavevectors}) and removes a
particle from either right or left Fermi point.  The two contributions
resulting from such treatment are
\begin{eqnarray}
  Z_{l,\uparrow}&=&\exp\left[-i\,\frac{\pi}{2}\,l
                      +\frac{i}{4}\,\varphi_L^{}(l)
                      +\frac{3i}{4}\,\varphi_R^{}(l)\right]
\nonumber\\
                  &&+\exp\left[i\,\frac{\pi}{2}\,l
                      -\frac{3i}{4}\,\varphi_L^{}(l)
                      -\frac{i}{4}\,\varphi_R^{}(l)\right].
  \label{eq:Z_lup}
\end{eqnarray}
Replacing the chiral bosonic fields in Eqs.~(\ref{eq:Z_ldown}) and
(\ref{eq:Z_lup}) with their non-chiral versions
(\ref{eq:non-chiral_bosons}), we find
\begin{equation}
    Z_{l,\gamma}=
        e^{i\frac{\pi}{2} l}
        e^{\mp i[\varphi(l)+\frac{1}{2}\vartheta(l)]}
        +e^{-i\frac{\pi}{2} l}
        e^{\pm i[\varphi(l)-\frac{1}{2}\vartheta(l)]},
\label{eq:Z_bosonization}
\end{equation}
where the upper and lower signs correspond to $\gamma=\,\uparrow$ and
$\downarrow$, respectively.  A generalization of the bosonization rule
(\ref{eq:Z_bosonization}) to the case of non-vanishing magnetization is
discussed in Appendix~\ref{sec:magnetization}.

\subsection{Two-step bosonization procedure for the electron operators}
\label{sec:two-step}

In this paper we consider one-dimensional electron systems with strong
repulsive interactions, when the spin exchange between electrons is
strongly suppressed, $J\ll E_F$.  In such systems the bosonization of
electron operators $\psi_\gamma(x)$ can be performed in two steps.  At
energies $|\varepsilon|$ below the Fermi energy $E_F$ the charge
excitations can be bosonized, and the fermion operators take the form
(\ref{eq:our_annihilation_operator_bosonized}).  This expression does not
assume a specific relation between $\varepsilon$ and $J$, so the spin
excitations are accounted for accurately by the operators $Z_{l,\gamma}$.
On the other hand, if $|\varepsilon|\ll J$, the spin excitations can also
be bosonized, Eq.~(\ref{eq:Z_bosonization}).  To compare the resulting
expression for the electron destruction operators with those used in the
standard bosonization procedure, Eq.~(\ref{eq:electron_bosonization}), we
substitute Eq.~(\ref{eq:Z_bosonization}) into
Eq.~(\ref{eq:our_annihilation_operator_bosonized}).  This substitution
results in 4 terms in the expression for $\psi_\gamma(x)$.  To identify the
annihilation operator for the right-moving electron
(\ref{eq:electron_bosonization_right}), we combine the first term in the
brackets in Eq.~(\ref{eq:our_annihilation_operator_bosonized}) with the
first term in the right-hand side of Eq.~(\ref{eq:Z_bosonization}).  This
yields
\begin{eqnarray}
  \psi_{R\gamma}(x)&=&\frac{e^{-i\theta(x)}}{\sqrt{2\pi\alpha}}\, 
                     e^{i[k_F^h x+\phi(x)]}
\nonumber\\
      &&\times
        e^{-i\frac{\pi}{2} l}
        e^{\pm i[\varphi(l)-\frac{1}{2}\vartheta(l)]}
        \Big|_{l=\frac{1}{\pi}[k_F^hx+\phi(x)]}
\nonumber\\
      &\simeq&
\frac{e^{\frac{i}{2}k_F^h x}}{\sqrt{2\pi\alpha}}\, e^{-i\theta(x)}
                     e^{\frac{i}{2}\phi(x)}
        e^{\pm i[\varphi(nx)-\frac{1}{2}\vartheta(nx)]}.
\label{eq:psi_right_two-step}
\end{eqnarray}
Expressing the bosonic fields via $\phi_{\rho,\sigma}$ and
$\theta_{\rho,\sigma}$ with the help of
Eqs.~(\ref{eq:relations_between_charge_variables}) and
(\ref{eq:relations_between_spin_variables}), we find that our result
(\ref{eq:psi_right_two-step}) is equivalent to the standard expression
(\ref{eq:electron_bosonization_right}).  Similarly, combining the second
term in the brackets in Eq.~(\ref{eq:our_annihilation_operator_bosonized})
with the second term in the right-hand side of
Eq.~(\ref{eq:Z_bosonization}), one reproduces the bosonized expression for
the annihilation operator (\ref{eq:electron_bosonization_left}) of
the left-moving electrons.

To understand the meaning of the remaining two contributions to
$\psi_\gamma(x)$, let us consider the momenta of the charge and spin
excitations in Eq.~(\ref{eq:psi_right_two-step}).  By destroying the
right-moving holon, we change the momentum of the system by $k_F^h=2k_F$.
Thus to obtain expression for annihilation operator of electron with
momentum near the right Fermi point $k_F$, we chose the left-moving
component of the operator $Z_{l,\gamma}$, which reduces the momentum
change by $k_F$.  By choosing the other component of $Z_{l,\gamma}$ we
increase the total momentum change to $3k_F$.  The physical meaning of
such process amounts to removing an electron from the right Fermi point
with simultaneous transfer of another electron from the right to the left
Fermi point.  In interacting electron systems such processes are possible,
but the resulting ``shadow band'' features tend to be weak.

\section{Green's functions}
\label{sec:greens_functions}

A number of important physical properties of one-dimensional electron
systems, such as the tunneling density of states and the spectral
functions, are expressed in terms of single-electron Green's functions,
\begin{subequations}
  \label{eq:Greens_functions_definition}
  \begin{eqnarray}
    \label{eq:G+}
    G^+_\gamma(x,t) &=& 
        \langle \psi_\gamma(x,t) \psi_\gamma^\dagger(0,0)\rangle,
\\
    \label{eq:G-}
    G^-_\gamma(x,t) &=& 
        \langle \psi_\gamma^\dagger(0,0) \psi_\gamma(x,t)\rangle.
  \end{eqnarray}
\end{subequations}
It is well known that in the limit $|t|\to\infty$ the Green's functions
show non-trivial power law behavior, with exponents depending on the
interaction strength.\cite{dzyaloshinskii} This behavior is easily
obtained\cite{giamarchi} in the bosonization approach based on
Eq.~(\ref{eq:electron_bosonization}).  In the case of strongly interacting
electrons these results are valid at $|t|\gg \hbar/J$ and adequately
describe the physics of the system at low energies $|\varepsilon|\ll J\ll
E_F$.  

In this paper we are primarily interested in the regime of intermediate
energies, $J\ll |\varepsilon|\ll E_F$.  To find the Green's functions in
this case, instead of Eq.~(\ref{eq:electron_bosonization}) one can use the
more general expression (\ref{eq:our_annihilation_operator_bosonized}).
We start by transforming the electron annihilation operator
(\ref{eq:our_annihilation_operator_bosonized}) to a more convenient form.

\subsection{Annihilation operator for strongly interacting electrons}
\label{sec:annihilation}

We first rewrite the lattice operators $Z_{l,\gamma}$ in
Eq.~(\ref{eq:our_annihilation_operator_bosonized}) in terms of their
Fourier components,
\begin{eqnarray}
  \label{eq:Fourier_to_momentum}
  Z_{l,\gamma} &=& \int_{-\pi}^\pi \frac{dq}{2\pi}\, z_\gamma(q)\, e^{iql},
\\
  \label{eq:Fourier_from_momentum}
  z_\gamma(q) &=& \sum_{l=-\infty}^\infty Z_{l,\gamma}\, e^{-iql}.
\end{eqnarray}
Straightforward substitution of Eq.~(\ref{eq:Fourier_to_momentum}) into
(\ref{eq:our_annihilation_operator_bosonized}) yields
\begin{eqnarray}
\hspace{-2em}
  \psi_{\gamma}(x)&=&\frac{e^{-i\theta(x)}}{\sqrt{2\pi\alpha}}\, 
                     \int_{-\pi}^\pi \frac{dq}{2\pi}\, z_\gamma(q)
\nonumber\\
                  &&\times
                     [e^{i(1+\frac{q}{\pi})[k_F^h x+\phi(x)]} 
                    + e^{i(-1+\frac{q}{\pi})[k_F^h x+\phi(x)]}].
\label{eq:annihilation_operator_first_take}
\end{eqnarray}
The two terms in the integrand correspond to removal of right- and
left-moving holons, respectively.

Expression (\ref{eq:annihilation_operator_first_take}) can be simplified
by noticing that $z_\gamma(q)$ is a $2\pi$-periodic function of $q$,
\begin{equation}
  \psi_{\gamma}(x)=\frac{e^{-i\theta(x)}}{\sqrt{2\pi\alpha}}\, 
                     \int_{-3\pi}^\pi \frac{dq}{2\pi}\, z_\gamma(q)\,
                     e^{i(1+\frac{q}{\pi})[k_F^h x+\phi(x)]}.
\label{eq:annihilation_operator_second_take}
\end{equation}
This presentation of the fermion operator is not entirely satisfactory,
because the limits of integration here and in
Eq.~(\ref{eq:Fourier_to_momentum}) were set rather arbitrarily.  Indeed,
the argument of the operator $z_\gamma(q)$ is the change of momentum of
the spin subsystem when an electron is removed.  Since spins in the
Hamiltonian (\ref{eq:Heisenberg}) are attached to lattice sites, the
momentum changes by $q$ or $q\pm2\pi$, $q\pm 4\pi$, etc.\ are equivalent.
This symmetry is lost in the bosonized expression
(\ref{eq:annihilation_operator_second_take}), but can be restored by
extending the limits of $q$-integration,
\begin{equation}
  \psi_{\gamma}(x)=\frac{e^{-i\theta(x)}}{\sqrt{2\pi\alpha}}\, 
                     \int_{-\infty}^\infty \frac{dq}{2\pi}\, z_\gamma(q)\,
                     e^{i(1+\frac{q}{\pi})[k_F^h x+\phi(x)]}.
\label{eq:annihilation_operator_final}
\end{equation}

The origin of the ambiguity in the definition of the fermion operators can
be traced back to the bosonization of the holon operators in
Eq.~(\ref{eq:our_annihilation_operator}).  Indeed, by definition
(\ref{eq:l(x)}), the operator $l(x)$ has only integer eigenvalues equal to
the number of electrons in the region of space from $-\infty$ to $x$,
whereas its bosonized expression $l(x)=\frac1\pi[k_F^h x +\phi(x)]$ does
not explicitly possess this property.  To enforce the discreteness of
charge in Eq.~(\ref{eq:our_annihilation_operator_bosonized}) one can
understand $Z_{l(x),\gamma}$ as
\begin{equation}
  \label{eq:Z_discrete}
  Z_{l,\gamma}\Big|_{l=\frac{1}{\pi}[k_F^hx+\phi(x)]}
  \to \sum_l Z_{l,\gamma}\,\delta(\textstyle{\frac{1}{\pi}}[k_F^hx+\phi(x)]-l).
\end{equation}
Then, upon the Fourier transformation (\ref{eq:Fourier_to_momentum}), one
recovers Eq.~(\ref{eq:annihilation_operator_final}).

A similar procedure of bosonization of fermions while preserving the
discreteness of their number was suggested by Haldane.\cite{haldane2}
Apart from the two terms corresponding to the right- and left-moving
fermions, the expression for the fermion operator contains multi-particle
contributions with wavevectors near $\pm3k_F$, $\pm5k_F$, etc.  In a
typical bosonization calculation these additional terms give much smaller
contributions than the leading ones.  Thus the difference between the
results obtained using the two bosonization schemes is smaller than the
accuracy of the bosonization approximation, and can be ignored.  In our
case, by using expression (\ref{eq:annihilation_operator_second_take})
instead of (\ref{eq:annihilation_operator_final}), one obtains small
spurious features in the spectral function originating from the
arbitrarily chosen integration limits.  Thus from now on we use
Eq.~(\ref{eq:annihilation_operator_final}).

\subsection{Green's functions at intermediate energies}
\label{sec:Greens_functions_intermediate}

At small $J$ the time evolution of the spin degrees of freedom is very
slow, with the typical time scales of order $\hbar/J$.  Therefore to find
the Green's functions describing physical phenomena at relatively high
energies $|\varepsilon|\gg J$, one can neglect the time-dependence of the
correlators of operators $Z_{l,\gamma}$.  Then the substitution of
Eq.~(\ref{eq:annihilation_operator_final}) into the definitions
(\ref{eq:Greens_functions_definition}) of the Green's functions yields
\begin{equation}
  \label{eq:G_pm}
  G_\gamma^\pm(x,t)=\frac{1}{2\pi\alpha}
                    \int_{-\infty}^\infty \frac{dq}{2\pi}\,
                    c_\gamma^\pm(q)\, 
                    e^{i(1+\frac{q}{\pi})k_F^h x}\,
                    g_q^\pm(x,t).
\end{equation}
Here the static correlators $c_\gamma^\pm(q)$ are defined by
\begin{subequations}
  \label{eq:c}
\begin{eqnarray}
  c_\gamma^+(q) &=& 
      \sum_l\langle Z_{l,\gamma}^{} Z_{0,\gamma}^\dagger\rangle\,
      e^{-iql},
  \label{eq:c^+}
\\
  c_\gamma^-(q) &=& 
      \sum_l\langle Z_{0,\gamma}^\dagger Z_{l,\gamma}^{}\rangle\,
  e^{-iql},
  \label{eq:c^-}
\end{eqnarray}
\end{subequations}
where $\langle\dots\rangle$ denotes thermal averaging over the equilibrium
states of the Hamiltonian (\ref{eq:Heisenberg}).  The time dependence is
contained in the correlators $g_q^\pm$, defined in terms of the charge
variables,
\begin{subequations}
  \label{eq:g_q}
  \begin{equation}
    \label{eq:g_q^+}
    g_q^+(x,t)
        =
          \langle
            e^{i[(1+\frac{q}{\pi})\phi(x,t)-\theta(x,t)]}\,
            e^{-i[(1+\frac{q}{\pi})\phi(0,0)-\theta(0,0)]}
          \rangle,
  \end{equation}
  \begin{equation}
    \label{eq:g_q^-}
    g_q^-(x,t)
        =
          \langle
            e^{-i[(1+\frac{q}{\pi})\phi(0,0)-\theta(0,0)]}\,
            e^{i[(1+\frac{q}{\pi})\phi(x,t)-\theta(x,t)]}
          \rangle.
  \end{equation}
\end{subequations}

These correlation functions are easily computed using the standard
techniques for averaging the exponentials of bosonic fields.\cite{giamarchi}
In the most important regime $|t|\ll \hbar/T$ one finds
\begin{equation}
  \label{eq:g_q^pm_result}
    g_q^\pm(x,t)=
       \left(
         \frac{\pm i\alpha}{x-v_\rho t \pm i\alpha}
       \right)^{\zeta^+_q}
       \left(
         \frac{\mp i\alpha}{x+v_\rho t \mp i\alpha}
       \right)^{\zeta^-_q}
\end{equation}
where
\begin{equation}
\zeta^\pm_{q}=
\frac{1}{4}\left[\sqrt{K}\left(1+\frac{q}{\pi}\right)
                 \pm\frac{1}{\sqrt{K}}\right]^2.
\label{eq:zeta_definition}
\end{equation}

The correlators $c_\gamma^\pm(q)$ are determined by the properties of the
Heisenberg spin chain (\ref{eq:Heisenberg}).  Unlike the holon correlators
$g_q^\pm(x,t)$, in general $c_\gamma^\pm(q)$ cannot be computed using the
bosonization approach.  We discuss their behavior in detail below.

\subsection{Static correlators $c_\gamma^\pm(q)$}
\label{sec:c^pm}

To find $c_\gamma^\pm(q)$ one has to perform the averaging in
Eqs.~(\ref{eq:c}) over the eigenstates of the Heisenberg Hamiltonian
(\ref{eq:Heisenberg}).  The results depend crucially on the relation
between the temperature $T$ and the exchange constant $J$.  At $T\sim J$
it can only be studied numerically.  Some analytical results can be found
in the cases of high and low temperatures.

\subsubsection{General mathematical properties of $c_\gamma^+(q)$ 
               and $c_\gamma^-(q)$}

\label{sec:general}

We start by establishing interesting relations between the functions
$c_\gamma^+(q)$ and $c_\gamma^-(q)$, which follow from their definition
(\ref{eq:c}) and do not depend on the temperature or the specific form
(\ref{eq:Heisenberg}) of the Hamiltonian of the spin chain.  We first
notice that the correlators $\langle Z_{l,\gamma}^{}
Z_{0,\gamma}^\dagger\rangle$ and $\langle Z_{0,\gamma}^\dagger
Z_{l,\gamma}^{}\rangle$ are real, and satisfy the following relations
\begin{eqnarray*}
  \langle Z_{0,\gamma}^{}Z_{0,\gamma}^\dagger\rangle&=&1,
\\
\langle Z_{l,\gamma}^{}Z_{0,\gamma}^\dagger\rangle
  &=&\langle Z_{-l,\gamma}^{}Z_{0,\gamma}^\dagger\rangle,
\\
\langle Z_{0,\gamma}^\dagger  Z_{l,\gamma}^{}\rangle
 &=& \langle Z_{0,\gamma}^\dagger  Z_{-l,\gamma}^{}\rangle.
\end{eqnarray*}
From the definitions (\ref{eq:c}) it then follows that the correlators
$c_\gamma^\pm(q)$ are real and even functions of $q$.  Also, the
definition (\ref{eq:c^+}) of $c_\gamma^+(q)$ can be rewritten as
\begin{equation}
  \label{eq:c^+_auxiliary}
  c^+_\gamma(q)=1+2\,{\rm Re}\sum_{l=1}^\infty
                \langle Z_{l,\gamma}^{}Z_{0,\gamma}^\dagger\rangle
                \,e^{-iql}.
\end{equation}
Furthermore, one can establish a simple relation between operators
$Z_{l,\gamma}^{}Z_{0,\gamma}^\dagger$ and $Z_{0,\gamma}^\dagger
Z_{l,\gamma}^{}$.  Both of them add and remove a site with spin $\gamma$
at different positions on the spin chain.  If a site is added first, the
numbering of all the subsequent sites is shifted by 1.  Thus
$Z_{l,\gamma}^{}Z_{0,\gamma}^\dagger=Z_{0,\gamma}^\dagger
Z_{l-1,\gamma}^{}$.  (Commutation of operators $Z_{l,\gamma}^{}$ is
discussed in more detail in Appendix~\ref{sec:anticommutation}.)  Then,
using Eq.~(\ref{eq:c^-}) one obtains
\[
 \langle Z_{l,\gamma}^{}Z_{0,\gamma}^\dagger\rangle
  =\int_{-\pi}^\pi \frac{dq}{2\pi}\, c_\gamma^-(q)\, e^{iq(l-1)}.
\]
Substituting this relation into Eq.~(\ref{eq:c^+_auxiliary}) one finds
\begin{equation}
  \label{eq:plus-to-minus}
  c_\gamma^+(q)=1+c_\gamma^-(q)\cos q
                +\int_{-\pi}^\pi \frac{dq'}{2\pi}\, 
                 \frac{\sin\frac{q'+q}{2}}{\sin\frac{q'-q}{2}}\,
                 c_\gamma^-(q').
\end{equation}
Similarly, one can express $c_\gamma^-(q)$ in terms of $c_\gamma^+(q)$,
\begin{equation}
  \label{eq:minus-to-plus}
  c_\gamma^-(q)=c_\gamma^+(q)\cos q
                +\int_{-\pi}^\pi \frac{dq'}{2\pi}\, 
                 \cot\frac{q-q'}{2}\,
                 \sin q'\,
                 c_\gamma^+(q').
\end{equation}
The integrals over $q'$ in Eqs.~(\ref{eq:plus-to-minus}) and
(\ref{eq:minus-to-plus}) should be understood as the principal value.

The relations (\ref{eq:plus-to-minus}) and (\ref{eq:minus-to-plus})
simplify considerably at  $q=0$ and $q=\pi$.  In these cases we find
\begin{subequations}
  \label{eq:c(0)-c(pi)}
  \begin{eqnarray}
    c_\gamma^+(0)-c_\gamma^-(0) &=&
                 1+\langle Z_{0,\gamma}^{\dagger} Z_{0,\gamma}^{}\rangle,
\\
    c_\gamma^+(\pi)+c_\gamma^-(\pi) &=&
                 1-\langle Z_{0,\gamma}^{\dagger} Z_{0,\gamma}^{}\rangle.
  \end{eqnarray}
\end{subequations}
The operator $Z_{0,\gamma}^{\dagger} Z_{0,\gamma}^{}$ destroys and then
recreates a site with a given spin $\gamma$.  Thus the average $\langle
Z_{0,\gamma}^{\dagger} Z_{0,\gamma}^{}\rangle$ is the probability to find
spin $\gamma$ at site $l=0$.  In the absence of magnetic field one expects
$\langle Z_{0,\gamma}^{\dagger} Z_{0,\gamma}^{}\rangle=\frac12$, resulting
in
\begin{subequations}
  \label{eq:SU2}
\begin{eqnarray}
  c_\gamma^+(0)-c_\gamma^-(0)&=&\frac32,
\\
  c_\gamma^+(\pi)+c_\gamma^-(\pi)&=&\frac12.
\end{eqnarray}
\end{subequations}

The relations (\ref{eq:plus-to-minus}) and (\ref{eq:minus-to-plus}) follow
from the definition of the correlators $c^\pm_\gamma(q)$, and are not
sensitive to the specific Hamiltonian (\ref{eq:Heisenberg}) of the spin
chain.  The only property of the Hamiltonian (\ref{eq:Heisenberg})
important for the relations (\ref{eq:c(0)-c(pi)}) is the spin symmetry.
At the same level of universality one can find the full dependences
$c^\pm_\gamma(q)$ in the high-temperature limit.

\subsubsection{High temperature $T\gg J$}

At high temperatures, correlators similar to $c_\gamma^\pm(q)$ have been
studied by Penc and Serhan.\cite{serhan} They noticed that at $J/T\to 0$ the
spins are completely uncorrelated, and the correlator $\langle
Z_{l,\gamma}^{} Z_{0,\gamma}^\dagger\rangle$ is simply the probability of
finding $|l|$ random spins at sites $0,1, \ldots l-1$ (or
$l,l+1,\ldots,-1$ for negative $l$) in a given state $\gamma$.  Thus $\langle
Z_{l,\gamma}^{} Z_{0,\gamma}^\dagger\rangle=1/2^{|l|}$ and, similarly,
$\langle Z_{0,\gamma}^\dagger Z_{l,\gamma}^{}\rangle=1/2^{|l|+1}$.
Substituting these expressions into Eq.~(\ref{eq:c}), one finds
\begin{equation}
  \label{eq:c_incoherent}
  c^+_\gamma(q)=2c^-_\gamma(q)=\frac{3}{5-4\cos q},
  \quad
  T\gg J.
\end{equation}
One can check explicitly that these results are consistent with the
relations (\ref{eq:plus-to-minus}) and (\ref{eq:minus-to-plus}).

\subsubsection{Zero temperature}
\label{sec:greens_functions_zero_temperature}

At zero temperature the correlators $c^+_\gamma(q)$ and $c^-_\gamma(q)$
carry non-trivial information about the spin correlations in the ground
state of the antiferromagnetic Heisenberg spin chain
(\ref{eq:Heisenberg}).  Although the model (\ref{eq:Heisenberg}) is
exactly solvable, no exact results are known for the correlators
$c^\pm_\gamma(q)$.  The quantity analogous to $c^-_\gamma(q)$ was first
studied by Sorella and Parola\cite{sorella} who used the results of
numerical diagonalization of spin chains of up to 22
sites.\cite{footnote2} Their results indicated that $c^-_\gamma(q)$ is
extremely small at $0<q<\pi/2$, whereas $c^-_\gamma(q)\sim1$.  They
interpreted this behavior as the effect of ``spinon pseudo Fermi
surface,'' which can be rephrased as follows.

Let us consider the ground state of the fermionized version
(\ref{eq:Heisenberg_fermionized}) of the Heisenberg model.  Ignoring the
interactions (\ref{eq:H^z}) between the fermions, one can picture the
ground state as shown in Fig.~\ref{fig:groundstate}(a).  The function
$c^-_\uparrow(q)$ is defined in terms of the correlators $\langle
Z_{0,\uparrow}^{\dagger} Z_{l,\uparrow}^{}\rangle$.  By removing a
spin-$\uparrow$ site, the operator $Z_{l,\uparrow}$ destroys a fermion.
Ignoring for the moment other aspects of $Z_{l,\uparrow}$, one concludes
that $c^-_\uparrow(q)$ should vanish at $-\pi/2<q<\pi/2$, as the fermion
states with those values of the wavevector are empty, see
Fig.~\ref{fig:groundstate}(a).  A similar argument for the correlator
$c^+_\uparrow(q)$ shows that it should vanish in the region of $q$-space
below the Fermi surface.  Noticing that in the absence of magnetic field
specific spin direction is unimportant, one concludes
\begin{subequations}
  \label{eq:spinon_Fermi_surface}
  \begin{eqnarray}
    c^+_\gamma(q)&=&0 \quad{\rm at}\ \frac{\pi}{2}<q<\frac{3\pi}{2}
  \label{eq:spinon_Fermi_surface_plus}
\\
    c^-_\gamma(q)&=&0 \quad{\rm at}\ -\frac{\pi}{2}<q<\frac{\pi}{2}
  \label{eq:spinon_Fermi_surface_minus}
  \end{eqnarray}
\end{subequations}
Conditions (\ref{eq:spinon_Fermi_surface}) refer to the values of $q$
between $-\pi/2$ and $3\pi/2$; outside of that region they can be inferred
using the $2\pi$-periodicity of the functions $c^\pm_\gamma(q)$.

The above picture neglects two important aspects of the problem: (i) the
fermions interact with each other, Eq.~(\ref{eq:H^z}), and (ii) the
operator $Z_{l,\uparrow}$ not only destroys a fermion, but also removes a
site from the spin chain.  There is no \emph{a priori} reason to expect
that the effects of these approximations should be small.  However, the
bosonization approach to calculation of $c^\pm_\uparrow(q)$ near the Fermi
points $q=\pm\pi/2$ (outlined below) shows that although each effect is
significant, they mostly compensate each other.  Numerically, this
compensation results\cite{sorella,penc1} in $c^-_\gamma(q)<0.01$ for
$|q|<\pi/2$.  Based on their numerical results, Sorella and
Parola\cite{sorella} conjectured that in an infinitely long spin chain
$c^-_\gamma(q)=0$ at $|q|<\pi/2$.  From this point of view, the
numerically small values of $c^-_\gamma(q)$ in this interval should be
viewed as a finite-size effect.

A more detailed study of the static correlators of the spin chain
(\ref{eq:Heisenberg}) was performed by Penc \emph{et al.}\cite{penc1} They
studied both correlators $c^+_\gamma(q)$ and $c^-_\gamma(q)$ (in slightly
different notations) and found that not only $c^-_\gamma(q)$ is
numerically small at $0<q<\pi/2$, but also $c^+_\gamma(q)$ is small at
$\pi/2<q<\pi$.  Although these observations are consistent with the
single-particle Fermi surface prediction (\ref{eq:spinon_Fermi_surface}),
no significant size effect was observed.  Penc \emph{et al.}\cite{penc1}
concluded that contrary to Eq.~(\ref{eq:spinon_Fermi_surface}),
correlators $c^+_\gamma(q)$ and $c^-_\gamma(q)$ do not vanish exactly in
the ``forbidden'' regions of the wavevector $q$.

\begin{figure}[t]
 \resizebox{.45\textwidth}{!}{\includegraphics{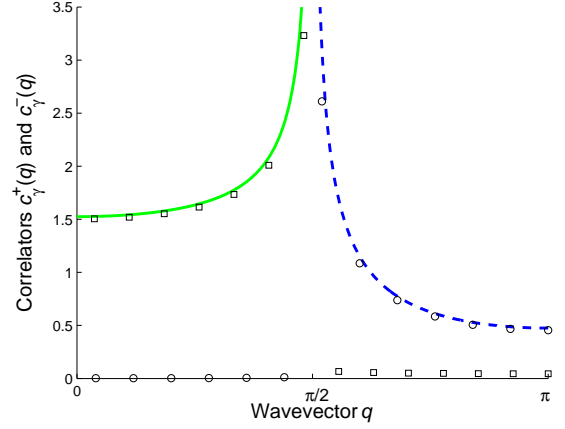}}
\caption{\label{fig:spinonFermi} The solution  of the integral equation
  (\ref{eq:plus-to-minus}) with conditions
  (\ref{eq:spinon_Fermi_surface}).  The solid and dashed lines show the
  behavior of $c^+_\gamma(q)$ and $c^-_\gamma(q)$ described by
  Eqs.~(\ref{eq:c^+muskhelishvili}) and (\ref{eq:c^-muskhelishvili}),
  respectively.  Squares and circles represent the results of numerical
  solution\cite{penc1,private} for the 26-site chain
  (\ref{eq:Heisenberg}).}
\end{figure}

We now show analytically that although the conditions
(\ref{eq:spinon_Fermi_surface}) may be a good approximation of actual
behavior of correlators $c^\pm_\gamma(q)$, they contradict to the
spin-rotation symmetry of the problem.  We showed in
Sec.~\ref{sec:general} that the correlators $c^+_\gamma(q)$ and
$c^-_\gamma(q)$ are related to each other, and if one of them is known on
the interval of $q$ of length $2\pi$, the other can be obtained using the
relations (\ref{eq:plus-to-minus}) and (\ref{eq:minus-to-plus}).
Conditions (\ref{eq:spinon_Fermi_surface}) define $c^+_\gamma(q)$ over
half the period $2\pi$ and $c^-_\gamma(q)$ over the other half.  This is
sufficient to uniquely define both functions at the remaining regions of
$q$-space.  To accomplish that one can solve the integral equation
(\ref{eq:minus-to-plus}) with conditions (\ref{eq:spinon_Fermi_surface}).
This mathematical problem belongs to the class of singular integral
equations, and can be solved by applying the well-known
techniques,\cite{muskhelishvili} see Appendix~\ref{sec:muskhelishvili}.
The solution has the form
\begin{subequations}
  \label{eq:c_muskhelishvili}
\begin{equation}
  \label{eq:c^+muskhelishvili}
  c^+_\gamma(q)=
       \frac{c_0}{\sqrt{\cos q}}
       \exp\left(
            -\frac{1}{\pi} 
             \int_{-\pi/2}^{\pi/2}
             \ln\left|\sin\frac{q-q'}{2}\right| dq'
           \right)
\end{equation}
for $-\pi/2<q<\pi/2$ and
\begin{equation}
  \label{eq:c^-muskhelishvili}
  c^-_\gamma(q)=
       \frac{c_0}{\sqrt{|\cos q|}}
       \exp\!\left(
            -\frac{1}{\pi} 
             \int_{-\pi/2}^{\pi/2}
             \ln\sin\frac{q-q'}{2}
             \,dq'
           \right)
\end{equation}
\end{subequations}
for $\pi/2<q<3\pi/2$.  The normalization constant $c_0$ is found using
Eq.~(\ref{eq:c(0)-c(pi)}),
\begin{equation}
  \label{eq:c_0}
   c_0=\frac{1}{2\cosh\frac{2\bm G}{\pi}},
\end{equation}
where ${\bm G}\approx0.91597$ is the Catalan's constant.  The solution
(\ref{eq:c^+muskhelishvili}), (\ref{eq:c^-muskhelishvili}) is plotted
in Fig.~\ref{fig:spinonFermi}.  

As expected, our solution gives an excellent approximation to the
numerical data of Ref.~\onlinecite{penc1}.  Nevertheless, the results
(\ref{eq:spinon_Fermi_surface}), (\ref{eq:c_muskhelishvili}) are not
exact.  Indeed, from Eqs.~(\ref{eq:c^+muskhelishvili}) and
(\ref{eq:c^-muskhelishvili}) we find
\begin{subequations}
\begin{eqnarray}
  c^+_\gamma(0)&=&\frac{2}{1+e^{-4{\bm G}/\pi}}\approx 1.525,
\\
  c^-_\gamma(\pi)&=&\frac{2}{e^{4{\bm G}/\pi}+1}\approx 0.475.
\end{eqnarray}
\end{subequations}
Taking into consideration that $c^+_\gamma(\pi)= c^-_\gamma(0)=0$,
Eq.~(\ref{eq:spinon_Fermi_surface}), we find that the spin-symmetry
conditions (\ref{eq:SU2}) are not satisfied exactly.

To gain better insight into the properties of the correlators
$c^\pm_\gamma(q)$, let us consider the case when the wavevector approaches
the Fermi point, $q\to\pi/2$.  In this regime one can apply the
bosonization approach of Sec.~\ref{sec:Z-bosonization}.  Using the
expressions (\ref{eq:Z_bosonization}) for the operators $Z_{l,\gamma}$ and
applying standard techniques\cite{giamarchi} for averaging the
exponentials of bosonic field over the ground state of the Hamiltonian
(\ref{eq:H_0}), we find
\begin{eqnarray}
  \label{eq:ZZdagger_average}
  \langle Z_{l,\gamma}^{} Z_{0,\gamma}^\dagger\rangle\
  &=& \left[
      e^{i\frac\pi2l}
      \left(\frac{\tilde\alpha}{\tilde\alpha+il}\right)^{1/2}
     +e^{-i\frac\pi2l}
      \left(\frac{\tilde\alpha}{\tilde\alpha-il}\right)^{1/2}
      \right]
\nonumber\\
  &&\times
      \left(
          \frac{\tilde\alpha^2}{\tilde\alpha^2+l^2}
      \right)^{\frac14\left(\mathcal{K}+\frac{1}{4\mathcal
            K}-1\right)}.
\end{eqnarray}
In the limit of low energies the Luttinger-liquid constant $\mathcal
K\to1/2$.  Using this value one can find the correlator $c^\pm(q)$ near
the Fermi point $q=\pi/2$ by substituting the component of the Green's
function (\ref{eq:ZZdagger_average}) given by the first term in the
brackets into Eq.~(\ref{eq:c^+}) and replacing the sum over $l$ with an
integral:
\begin{equation}
  \label{eq:c^+bosonization_v1}
  c^+_\gamma(q) = \int_{-\infty}^{\infty} 
                 \left(\frac{\tilde\alpha}{\tilde\alpha+il}\right)^{1/2}
                 e^{-i(q-\pi/2)l} dl.
\end{equation}
It is important to note that the integrand is analytic in the lower
complex half-plane.  Thus $c^+_\gamma(q)=0$ for $q>\pi/2$.  This
conclusion agrees with the prediction of the spinon Fermi surface
approximation (\ref{eq:spinon_Fermi_surface}).  At $q<\pi/2$ the
integration in Eq.~(\ref{eq:c^+bosonization_v1}) is straightforward, and
we find
\begin{subequations}
  \label{eq:c_bosonization_result}
\begin{equation}
  \label{eq:c^+bosonization_result}
  c^+_\gamma(q) =
  \chi\frac{\Theta\big(\frac{\pi}{2}-q\big)}
           {\sqrt{\frac{\pi}{2}-q}},                 
\end{equation}
where $\Theta(x)$ is the unit step function, and the value of the
numerical coefficient $\chi\sim1$ cannot be determined within the
bosonization approach.  A similar calculation for $c^-_\gamma(q)$ results
in
\begin{equation}
  \label{eq:c^-bosonization_result}
  c^-_\gamma(q) =
  \chi\frac{\Theta\big(q-\frac{\pi}{2}\big)}
           {\sqrt{q-\frac{\pi}{2}}},                 
\end{equation}
\end{subequations}
also in agreement with the prediction (\ref{eq:spinon_Fermi_surface}) of
the spinon Fermi surface picture.  

The inverse-square-root singularities (\ref{eq:c_bosonization_result}) of
the correlators $c^\pm_\gamma(q)$ at the Fermi point are consistent with
the numerical data of Ref.~\onlinecite{penc1} and our expressions
(\ref{eq:c_muskhelishvili}), see also Fig.~\ref{fig:spinonFermi}.  In
particular, the asymptotes of the expressions (\ref{eq:c_muskhelishvili})
at $q\simeq \pi/2$ are given by Eq.~(\ref{eq:c_bosonization_result}) with
$\chi\approx 0.85$.

It is important to note that the results (\ref{eq:c_bosonization_result})
of the bosonization treatment of the static correlators $c^\pm_\gamma(q)$
do not prove the validity of the spinon Fermi surface picture.  Indeed,
the bosonized expressions (\ref{eq:Z_bosonization}) for operators
$Z_{l,\gamma}$ are only applicable asymptotically near the Fermi point,
$q\to\pm\pi/2$.  Consequently, the presence of the step function
$\Theta$ in the results (\ref{eq:c_bosonization_result}) should be
interpreted as
\begin{equation}
  \label{eq:c/c}
   \lim_{\delta \to +0}
   \frac{c^+_\gamma(\frac\pi2+\delta)}{c^+_\gamma(\frac\pi2-\delta)}=
   \lim_{\delta \to +0}
   \frac{c^-_\gamma(\frac\pi2-\delta)}{c^-_\gamma(\frac\pi2+\delta)}=0.  
\end{equation}
Let us now show that the correlator $c^+_\gamma(q)$  not only does not vanish
at $q>\pi/2$, but in fact diverges at $q\to \pi/2+0$.

We start by noticing that the conclusion $c^+_\gamma(q)=0$ at $q>\pi/2$
holds only for $\mathcal{K}=1/2$.  Indeed, at $\mathcal{K}\neq1/2$ the
last factor in the expression (\ref{eq:ZZdagger_average}) is no longer
unity.  More importantly, it is no longer analytic in the lower complex
half-plane, leading to the non-vanishing value of the integral
(\ref{eq:c^+bosonization_v1}) at $q>\pi/2$.

The low-energy properties of the Heisenberg spin chain
(\ref{eq:Heisenberg}) are adequately described by the sine-Gordon
Hamiltonian (\ref{eq:Heisenberg_bosonized}).  The parameters $\mathcal
K$ and $g$ renormalize at low energies or, equivalently, long length
scales $L$ as
\begin{eqnarray}
  \mathcal{K} &=& \frac12 +\frac{y}{4},
\\
  g&=&\pi\tilde v y,
\\
  y&=&\frac{1}{\ln L}.
  \label{eq:sine-Gordon_scaling}
\end{eqnarray}
Thus at low energies $\mathcal{K}$ deviates slightly from 1/2.  This
deviation gives a correction to the correlator
(\ref{eq:c^+bosonization_v1}) which can be obtained by expansion of
Eq.~(\ref{eq:ZZdagger_average}) in powers of $y=4(\mathcal{K} -1/2)$,
\begin{equation}
  \label{eq:delta_c^+bosonization_v1}
  \delta c^+_\gamma(q) = -\frac{y^2}{32}
                 \int_{-\infty}^{\infty} 
                 \left(\frac{\tilde\alpha}{\tilde\alpha+il}\right)^{1/2}
                 \ln\frac{\tilde\alpha^2+l^2}{\tilde\alpha^2}\,
                 e^{-i(q-\pi/2)l} dl.
\end{equation}
At $q<\pi/2$ the correction (\ref{eq:delta_c^+bosonization_v1}) is small
compared to the leading term (\ref{eq:c^+bosonization_result}) and can be
ignored.  However, unlike Eq.~(\ref{eq:c^+bosonization_result}), the
correction (\ref{eq:delta_c^+bosonization_v1}) does not vanish  at
$q>\pi/2$,
\begin{equation}
  \label{eq:delta_c^+bosonization_result}
  \delta c^+_\gamma(q) =
  \frac{\pi}{32}\, y^2
  \frac{\chi}{\sqrt{q-\frac{\pi}{2}}},                 
\end{equation}
where $\chi$ is the same numerical coefficient as in
Eq.~(\ref{eq:c_bosonization_result}).

The deviation of the Luttinger liquid parameter $\mathcal{K}$ from the
limiting value of 1/2 is not the only source of corrections to
$c^+_\gamma(q)$.  Additional corrections originate from the small
sine-Gordon term (\ref{eq:V}).  One can account for this term to second
order perturbation theory using the standard techniques.\cite{giamarchi}
The resulting correction is factor of 2 greater than
Eq.~(\ref{eq:delta_c^+bosonization_result}).  [Apart from direct
calculation, this can be shown to follow from the spin-rotation symmetry
of the problem, see Appendix~\ref{sec:g-ology}.\label{sec:c-correction}]
We therefore conclude that at $q$ slightly above the Fermi point $\pi/2$
the correlator $c^+_\gamma$ is given by
\begin{equation}
  \label{eq:c^+correction}
  c^+_\gamma(q) =
  \frac{3\pi}{32}\, y^2
  \frac{\chi}{\sqrt{q-\frac{\pi}{2}}},
\quad
  q\to \frac\pi2+0.
\end{equation}

The above calculation was performed to second-order perturbation theory in
coupling constant $y$ and did not account for its scaling
(\ref{eq:sine-Gordon_scaling}).  Since the scaling is logarithmic, and
thus slow compared to the leading power-law behavior in
Eq.~(\ref{eq:c^+correction}), one can simply substitute the expression
(\ref{eq:sine-Gordon_scaling}) into (\ref{eq:c^+correction}) choosing the
proper value of the length scale $L\sim1/(q-\pi/2)$.  We therefore
conclude that near the Fermi point $q=\pi/2$ the correlator $c^+_\gamma(q)$
behaves as
\begin{subequations}
  \label{eq:c_sine-Gordon}
\begin{equation}
  \label{eq:c^+sine-Gordon}
  c^+_\gamma(q)=\left\{
  \begin{array}[c]{ll}
    \frac{\chi}{\sqrt{\frac{\pi}{2}-q}}, & q\to \frac\pi2-0,\\[3ex]
    \frac{3\pi}{32}\, \frac{1}{\ln^2(q-\frac{\pi}{2})}
  \frac{\chi}{\sqrt{q-\frac{\pi}{2}}}, & q\to \frac\pi2+0.
  \end{array}
  \right.
\end{equation}
Analogous calculation for the correlator $c^-_\gamma$ gives
\begin{equation}
  \label{eq:c^-sine-Gordon}
  c^-_\gamma(q)=\left\{
  \begin{array}[c]{ll}
    \frac{\chi}{\sqrt{\frac{q-\pi}{2}}}, & q\to \frac\pi2+0,\\[3ex]
    \frac{3\pi}{32}\, \frac{1}{\ln^2(\frac{\pi}{2}-q)}
  \frac{\chi}{\sqrt{\frac{\pi}{2}-q}}, & q\to \frac\pi2-0.
  \end{array}
  \right.
\end{equation}
\end{subequations}

Our results (\ref{eq:c_sine-Gordon}) reaffirm our earlier observation that
the correlators $c^\pm_\gamma(q)$ do not vanish exactly on one side of the
Fermi point, as expected from the spinon Fermi surface picture,
cf.~Eq.~(\ref{eq:spinon_Fermi_surface}).  Instead the correlators diverge
at $q\to\pi/2$, albeit slower than on the ``main'' side of the Fermi
point.  The numerical data of Ref.~\onlinecite{penc1} shown in
Fig.~\ref{fig:spinonFermi} indicate that the small values of
$c^+_\gamma(q)$ at $q>\pi/2$ do increase near the Fermi point.  However,
studies of much longer systems are needed to verify the asymptotes
(\ref{eq:c_sine-Gordon}).

\section{Spectral functions}
\label{sec:spectral_functions}

The spectral functions $A^+_\gamma(k,\omega)$ and $A^-_\gamma(k,\omega)$
of a one-dimensional electron system are defined as Fourier transforms
\begin{equation}
  A^\pm_\gamma(k,\omega) =
    \int^\infty_{-\infty}dx\int^\infty_{-\infty}\frac{dt}{2\pi}\,
    e^{-ikx+i\omega t} G^\pm_\gamma(x,t)
  \label{eq:spectral_functions_definition}
\end{equation}
of the Green's functions (\ref{eq:Greens_functions_definition}).  The
components $A^+_\gamma(k,\omega)$ and $A^-_\gamma(k,\omega)$ characterize
the particle and hole parts of the excitation spectrum, respectively.  In
this section we study the behavior of the spectral functions at
frequencies $\omega$ in the range
\begin{equation}
  \label{eq:frequency_range}
  J, T\ll \hbar\omega\ll D_\rho.
\end{equation}
Under these restrictions one can express the Green's functions in the form
(\ref{eq:G_pm}), (\ref{eq:c}), (\ref{eq:g_q^pm_result}).  Performing the
integration with respect to $x$ and $t$, one finds
\begin{eqnarray}
\hspace{-2em}
A_{\gamma}^\pm(k,\omega)&=&
\int_{q_k-q_\omega}^{q_k+q_\omega}dq\frac{\Theta(\pm\omega)}{2v_\rho k_F^h}
     \frac{(\alpha k_F^h/2\pi)^{\zeta^+_q+\zeta^-_q-1}}
          {\Gamma(\zeta^+_q)\Gamma(\zeta^-_q)}
\nonumber\\
&&\times
\frac{c^\pm_\gamma(q)}
     {[q-q_k+q_\omega]^{1-\zeta^\mp_q}[q_k+q_\omega-q]^{1-\zeta^\pm_q}},
\label{eq:A_general}
\end{eqnarray}
where
\begin{equation}
  \label{eq:q_k_definition}
  q_k = \frac{\pi}{k_F^h}(k-k_F^h),
  \quad
  q_\omega=\frac{\pi|\omega|}{v_\rho k_F^h}.
\end{equation}
According to Eq.~(\ref{eq:A_general}) at small frequencies $\omega\ll
v_\rho k_F^h$ the spinon wavevector $q$ must be close to $q_k$.  This can
be understood in terms of the energy and momentum conservation laws.
Since the energy $\omega$ is small, the holon component of the electron
must have momentum near $\pm k_F^h$.  If the electron momentum $k$ is near
one of those values, the spinons carry no momentum, i.e., $q=0$ (or
$2\pi$), in agreement with $q\approx q_k\approx0$.  On the other hand, if
$k$ is not near $\pm k_F^h$, the difference of the momenta $k-k_F^h=nq_k$
is transferred into the spin subsystem.  (Note that the electron density
$n=k_F^h/\pi$.)

\subsection{Zero momentum peak}
\label{sec:peak_k=0}

To find the momentum dependence of the spectral functions
(\ref{eq:A_general}) at low frequencies, one can replace $q\to q_k$ in
$c^\pm_\gamma(q)$ and $\zeta^\pm_q$.  The remaining integration is
straightforward and gives
\begin{equation}
  \label{eq:A^pm_result}
  A^\pm_\gamma(k,\omega) =
  \frac{\Theta(\pm\omega)c_\gamma^\pm(q_k)}{2v_\rho k_F^h
                         \Gamma\big(\zeta(k)\big)}
\left(\frac{D_\rho}{\hbar|\omega|}\right)^{1-\zeta(k)},
\end{equation}
where 
\begin{equation}
  \label{eq:zeta(k)}
  \zeta(k)=\zeta^+_{q_k^{}}+\zeta^-_{q_k^{}}
          =\frac{K}{2}  \left(\frac{k}{k_F^h}\right)^2 + \frac{1}{2K}.
\end{equation}
In the low-frequency limit $\hbar|\omega|/D_\rho\to0$ the spectral
function $A^\pm_\gamma$ has a sharp Gaussian peak as a function of the
electron momentum centered at $k=0$.  In the spin-incoherent limit
$J/T\to0$ a similar Gaussian peak in the momentum dependence of the
spectral function was found by Fiete \textit{et al.}\cite{fiete3} Our
expression (\ref{eq:A^pm_result}) is valid at arbitrary $J/T$, with the
temperature dependence entering Eq.~(\ref{eq:A^pm_result}) via the
functions $c^\pm_\gamma(q)$.  In particular, using the numerical
results\cite{penc1} for $c^\pm_\gamma(q)$ we can access the
zero-temprature limit.

The peak (\ref{eq:A^pm_result}) gives the leading contribution to the
density of states at low energies
\begin{eqnarray}
  \label{eq:nu}
\hspace{-1.8em}
    \nu_\gamma^\pm(\varepsilon) 
       &=& 
       \int_{-\infty}^{\infty}\frac{dk}{2\pi\hbar} 
                               A^\pm_\gamma(k,\varepsilon/\hbar)
\nonumber\\
       &=&\frac{\Theta(\pm\varepsilon)}{\pi\hbar v_\rho}
       \sqrt{\frac{\pi}{8 K}}
       \frac{c_\gamma^\pm(\pi)}{\Gamma\left(\frac{1}{2K}\right)}\!
       \left(\frac{D_\rho}{|\varepsilon|}\right)^{1-\frac{1}{2K}}\!
       \frac{1}{\sqrt{\ln\frac{D_\rho}{|\varepsilon|}}}.
\end{eqnarray}
This result reproduces the expressions for the tunneling density of states
at $\varepsilon/D_\rho\to0$ reported in Ref.~\onlinecite{brief}.

It is important to note that the Gaussian peak in the spectral function
(\ref{eq:A^pm_result}) cannot be obtained within the standard
Luttinger-liquid theory of spectral functions,\cite{meden,voit} which
applies only near $k=\pm k_F$.  The peak (\ref{eq:A^pm_result}) is due to
the holon states with wavevectors near $k_F^h=2k_F$, and the spin
excitations with $q$ near $\pi$, well below the spinon Fermi surface, see
Fig.~\ref{fig:groundstate}.

An interesting consequence of the fact that the Gaussian peak is dominated
by spin excitations away from the Fermi points is the dramatic difference
of the peak heights for the spectral functions $A_\gamma^+$ and
$A_\gamma^-$.  Indeed, as we discussed in Sec.~\ref{sec:c^pm}, at zero
temperature $c_\gamma^+(\pi)\ll c_\gamma^-(\pi)$, whereas at $T\gg J$ one
has $c_\gamma^+=2 c_\gamma^-$.  Thus the heights of the peaks at $k=0$
depend strongly on temperature, and the peak in $A_\gamma^-$ is much more
pronounced than that in $A_\gamma^+$ at $T\ll J$.

In our derivation of the expression (\ref{eq:A^pm_result}) for the
spectral functions we replaced $q\to q_k$ in the arguments of the spin
correlators $c_\gamma^\pm(q)$ in Eq.~(\ref{eq:A_general}).  This procedure
is well justified if $q_k$ is not too close to points $\pi(2s-1)/2$ (with
$s=0,\pm1,\pm2,\ldots$), where at zero temperature the correlators
$c_\gamma^\pm(q)$ have sharp inverse-square-root singularities, see
Fig.~\ref{fig:spinonFermi}.  In addition, one can still use
Eq.~(\ref{eq:A^pm_result}) if the singularities of $c_\gamma^\pm(q)$ are
smeared by finite temperature by $\delta q\sim T/J\gg q_\omega$.  However,
at very low temperatures
\begin{equation}
  \label{eq:low_temperature_condition}
  T\lesssim\frac{\hbar\omega J}{D_\rho}
\end{equation}
the approximation leading to Eq.~(\ref{eq:A^pm_result}) fails near
$q_k=\pi(2s-1)/2$.  Thus at low temperatures the spectral functions have
non-trivial behavior in the vicinity of $k=(2s+1)k_F$.  Below we consider
the zero-temperature behavior of the spectral functions near these
points. 

\subsection{Fermi surface features}
\label{sec:fermi_surface_features}

When the electron wavevector approaches the Fermi point $k_F$ we have
$q_k\to-\pi/2$.  From Eqs.~(\ref{eq:c^+bosonization_result}) and the
symmetry $c_\gamma^+(q)=c_\gamma^+(-q)$ we conclude that near $q=-\pi/2$
one can approximate $c_\gamma^+(q)$ as
\begin{equation}
  \label{eq:c^+approximated}
  c^+_\gamma(q) =
  \chi\frac{\Theta\big(\frac{\pi}{2}+q\big)}
           {\sqrt{\frac{\pi}{2}+q}}.
\end{equation}
To explore the fine structure of the Fermi surface features at zero
temperature, instead of the approximate expression (\ref{eq:A^pm_result})
one should use the more general formula (\ref{eq:A_general}).  Then
instead of one singularity at $q_k=-\pi/2$ one finds two singularities
at $q_k= -\pi/2\pm q_\omega$.  Given the definitions
(\ref{eq:q_k_definition}), one concludes that the spectral function
$A_\gamma^+$ has singularities at two points $k=k_F\pm\omega/v_\rho$,
above and below $k_F$.

In evaluating the singular terms in the spectral function
(\ref{eq:A_general}) one can approximate the exponents $\zeta^\pm_q$ by
their values at $q=-\pi/2$.  Using the expressions
(\ref{eq:zeta_definition}) we then find
$\zeta^-_{-\pi/2}=\zeta^+_{-\pi/2}-1/2=\zeta_0$, where
\begin{equation}
  \label{eq:zeta_rho}
  \zeta_0 = \frac{(2-K)^2}{16K}=\frac{(1-K_\rho)^2}{8K_\rho}.
\end{equation}

The stronger of the two singularities is the one above $k_F$, at
$k=k_F+\omega/v_\rho$.  It appears when the inverse square root
singularity (\ref{eq:c^+approximated}) in $c_\gamma^+(q)$ is near the
lower limit of the integral (\ref{eq:A_general}).  The most singular term
in the spectral function $A_\gamma^+$ can then be found by extending the
upper limit to infinity, resulting in
\begin{eqnarray}
   A_{\gamma}^+(k,\omega)&\simeq&
    \frac{\chi(\alpha/2v_\rho)^{2\zeta_0-1/2}B}
         {2\pi\Gamma\big(\zeta_0+\frac12\big)\Gamma(\zeta_0)}
\sqrt{\frac{\pi}{2v_\rho k_F}}
\nonumber\\
&&\times
[2\omega|\omega-v_\rho(k-k_F)|]^{\zeta_0 -\frac{1}{2}},
\label{eq:A_singularity}
\end{eqnarray}
where the constant $B$ is defined in terms of the beta function
\begin{equation}
  \label{eq:B}
  B=\left\{
  \begin{array}[c]{ll}
  B\big({\textstyle\frac{1}{2}-\zeta_0,\frac{1}{2}}\big), 
            & \omega-v_\rho(k-k_F)\to+0,\\[2ex]
  B\big(\textstyle{\frac{1}{2}-\zeta_0,\zeta_0}\big), 
            & \omega-v_\rho(k-k_F)\to-0.
  \end{array}
  \right.
\end{equation}
The power-law singularity (\ref{eq:A_singularity}) of the spectral
function $A^+_\gamma$ at $k=k_F+\omega/v_\rho$ with the exponent
$\zeta_0-1/2$ is consistent with the results of the Luttinger-liquid
theory.\cite{meden,voit}  

It is worth noting that at
\begin{equation}
  \label{eq:K-condition}
  K=2K_\rho>6-4\sqrt2\approx0.343
\end{equation}
the spectral function (\ref{eq:A_singularity}) diverges at
$k=k_F+\omega/v_\rho$.  Nevertheless, the dominant contribution to the
density of states (\ref{eq:nu}) at low energies is given by the Gaussian
peak (\ref{eq:A^pm_result}) at $k=0$.

The second singularity of the spectral function $A_\gamma^+$ is below
$k_F$, at $k=k_F-\omega/v_\rho$.  It emerges when the singularity
(\ref{eq:c^+approximated}) of $c_\gamma^+(q)$ is at the upper limit of the
integral (\ref{eq:A_general}).  As $k$ approaches $k_F-\omega/v_\rho$ from
above, the width of the integration regions shrinks to zero, and at
$k<k_F-\omega/v_\rho$ the integral vanishes.  The singularity at $k\to
k_F-\omega/v_\rho$ has the form
\begin{eqnarray}
   A_{\gamma}^+(k,\omega)&\simeq&
    \frac{\Theta\big(k-(k_F-\omega/v_\rho)\big)
            \chi\zeta_0(\alpha/2v_\rho)^{2\zeta_0-1/2}}
         {2\pi\big[\Gamma\big(\zeta_0+\frac12\big)\big]^2\sqrt{2v_\rho k_F}}
\nonumber\\
&&\times
(2\omega)^{\zeta_0 -1} [\omega+v_\rho(k-k_F)]^{\zeta_0}.
\label{eq:A_secondary_singularity}
\end{eqnarray}
Unlike the feature (\ref{eq:A_singularity}) above $k_F$, this term always
vanishes at the singularity because $\zeta_0>0$.  The power-law feature
in the spectral function with the exponent $\zeta_0$ was obtained
earlier\cite{meden,voit} in the framework of the Luttinger liquid theory.

The fact that the power-law feature (\ref{eq:A_secondary_singularity})
appears only on one side of the point $k=k_F-\omega/v_\rho$ is a
consequence of the ``spinon Fermi surface'' approximation.  In a more
careful treatment, the correlator $c_\gamma^+(q)$ does not vanish at
$q<-\pi/2$, but instead has an inverse-square-root singularity with an
additional factor $1/\ln^2(q+\pi/2)$, cf. Eq.~(\ref{eq:c^+sine-Gordon}).
Thus the cusp (\ref{eq:A_secondary_singularity}) should appear on both
sides of the point $k=k_F-\omega/v_\rho$, albeit with an additional
logarithmic suppression factor at $k<k_F-\omega/v_\rho$.  In the Luttinger
liquid theory this feature would be caused by the marginally irrelevant
perturbation $\cos(2\sqrt2\,\phi_\sigma)$ that was not accounted for in
Refs.~\onlinecite{meden,voit}.

At $\omega<0$ the spectral function $A_\gamma^-$ has similar power-law
singularities at $k=k_F\pm\omega/v_\rho$.  The respective expressions can
be obtained by replacing $\omega\to-\omega$ and $k-k_F\to k_F-k$ in
Eqs.~(\ref{eq:A_singularity}) and (\ref{eq:A_secondary_singularity}).

\subsection{Shadow band features}
\label{sec:3k_F_singularities}

The singularities in the spectral functions emerge when excitations in
both charge and spin subsystems are near the Fermi points $k_F^h$ and
$(2s-1)\pi/2$, respectively.  In particular, if the holon momentum is
$k_F^h=2k_F$ and the spinon momentum $nq=(2k_F/\pi)(\pi/2)=k_F$, the
electron momentum is $3k_F$.  Formally, the features in the spectral
functions are caused by the singularities (\ref{eq:c_bosonization_result})
appearing inside the narrow integration region in
Eq.~(\ref{eq:A_general}).

Similarly to the singularities near $k_F$, one expects to find two
features, at $k=3k_F\pm\omega/v_\rho$, when either of the limits of
integration crosses the point $q=\pi/2$.  Since the features in the
spectral functions near $3k_F$ are weaker than the ones near $k_F$, we
discuss only the stronger of the two singularities in $A_\gamma^+$.  It
appears when the lower limit $q_k-q_\omega$ is near $\pi/2$ and
corresponds to $k=3k_F+\omega/v_\rho$.

The behavior of the spectral function is controlled by the exponents
(\ref{eq:zeta_definition}) at $q=\pi/2$, which can be expressed in terms
of a single parameter
\begin{equation}
  \label{eq:tilde_zeta_rho}
  \zeta_1= \frac{(3K-2)^2}{16K}=\frac{(3K_\rho-1)^2}{8K_\rho}.
\end{equation}
as $\zeta_q^- = \zeta_q^+ - 3/2 =\zeta_1$.  Then the singular
term in the spectral function takes the form
\begin{eqnarray}
\hspace{-1.5em}
   A_{\gamma}^+(k,\omega)&\simeq&
    \frac{\Theta\big(3k_F+\omega/v_\rho-k\big)
            \chi(\alpha/2v_\rho)^{2\zeta_1+1/2}}
         {2\pi\big(\zeta_1+\frac12\big)
         [\Gamma(\zeta_1)]^2
          \sqrt{2v_\rho k_F}}
\nonumber\\
&&\times
(2\omega)^{\zeta_1 +1/2} 
[\omega-v_\rho(k-3k_F)]^{\zeta_1-1/2}.
\label{eq:A_3kF_singularity}
\end{eqnarray}
Similarly to the singularity (\ref{eq:A_secondary_singularity}), one will
find a weaker feature on the other side of the singularity, i.e., at
$k\to(3k_F+\omega/v_\rho)+0$, if instead of
Eq.~(\ref{eq:c^+bosonization_result}) a more accurate approximation
(\ref{eq:c^+sine-Gordon}) for the correlator $c^+_\gamma(q)$ is applied.

The features in the spectral functions at $k=3k_F+\omega/v_\rho$ have been
observed in numerical data for the infinite-$U$ Hubbard model by Penc
\emph{et al.,}\cite{penc2} who identified it with the so-called shadow
band.\cite{kampf} Our formula (\ref{eq:A_3kF_singularity}) provides
analytic expression for the spectral function at the shadow band position
in the limit of low frequencies $\omega\ll D_\rho/\hbar$.  Unlike the
numerical treatment of Ref.~\onlinecite{penc2}, our result is not limited
to the Hubbard model with only on-site repulsion and can be applied to
systems with any interaction range.

In addition to the shadow band feature near $k=3k_F$, the periodicity of
the correlators $c^\pm_\gamma(q)$ results in singularities of the spectral
functions at all odd multiples of $k_F$.  Similar to the features near $k_F$
and $3k_F$, one finds a pair of singularities at
$k=(2s+1)k_F\pm\omega/v_\rho$ for $s=2,3,\ldots$.  The stronger
singularity in each pair is the one at $k=(2s+1)k_F+\omega/v_\rho$, where
one finds
\begin{eqnarray}
\hspace{-1.5em}
   A_{\gamma}^\pm(k,\omega)&\propto&
         |\omega|^{\zeta_s + s - 1/2} 
\nonumber\\
      &&\times
         \{\omega-v_\rho[k-(2s+1)k_F]\}^{\zeta_s-1/2},
\label{eq:A_(2s+1)kF_singularity}
\end{eqnarray}
with
\begin{equation}
  \label{eq:zeta_s}
  \zeta_s= \frac{[(2s+1)K-2]^2}{16K}=\frac{[(2s+1)K_\rho-1]^2}{8K_\rho}.
\end{equation}
At $s=0, 1$ this expression is consistent with our earlier results
(\ref{eq:A_singularity}) and (\ref{eq:A_3kF_singularity}).  For a given
$s$ the strongest (inverse square-root) divergence of the spectral
functions at $k=(2s+1)k_F+\omega/v_\rho$ is achieved when $\zeta_s=0$,
i.e., $K_\rho=1/(2s+1)$.  Since in the case of strong repulsion
$K_\rho<1/2$, this condition cannot be satisfied for $s=0$; the lowest
possible value of $\zeta_0$ is 1/16.

\section{Summary and discussion of the results}
\label{sec:discussion}

In this paper we have developed the theory of one-dimensional electron
systems in the regime of very strong interactions.  This regime emerges
when the repulsion between electrons strongly suppresses exchange of their
spins, $J\ll E_F$.  Our theory is based upon the Hamiltonian
$H=H_\rho+H_\sigma$, with the charge part
(\ref{eq:original_holon_Hamiltonian}) brought to the form
(\ref{eq:bosonized_holon_Hamiltonian}) by means of conventional
bosonization while the spin contribution $H_\sigma$ is the Hamiltonian of
the Heisenberg spin chain (\ref{eq:Heisenberg}).  The most important
ingredient of the theory is the expression (\ref{eq:our_operators}) for
the electron creation and annihilation operators in terms of the charge
and spin degrees of freedom.

In our technique the charge excitations are bosonized, and thus the
applicability of the results is limited to energies well below $E_F$.  On
the other hand, the spin excitations are treated more carefully, so we can
access the energy scales both below and above $J$.  At energies below $J$
the standard approach based on the bosonization procedure
(\ref{eq:electron_bosonization}) and the Hamiltonian
(\ref{eq:standard_bosonization}) can be applied.  We showed in
Sec.~\ref{sec:spin_bosonization} that at $\varepsilon\ll J$ our
expressions (\ref{eq:our_operators}) for the electron operators reproduce
the bosonization formulas (\ref{eq:electron_bosonization}).  The advantage
of our method is that unlike the bosonization procedure
(\ref{eq:electron_bosonization}) it can also be applied at energy scales
$\varepsilon \gtrsim J$.

The main difficulty in applying our technique is the need to find the
Green's functions of operators $Z_{l,\gamma}$ with the Heisenberg
Hamiltonian (\ref{eq:Heisenberg}).  The problem is simplified for the most
interesting case $\varepsilon\gg J$, when the slow time dependence of the
spin degrees of freedom can be ignored.  In this case the single-particle
Green's functions of electrons can be expressed in terms of the static
spin correlators $c_\gamma^\pm(q)$, which were studied in
Refs.~\onlinecite{sorella,penc1,penc3,serhan}.  Additional useful
properties of these correlators are derived in Sec.~\ref{sec:c^pm}.

We have applied our technique to the calculation of the spectral functions
of strongly interacting one-dimensional electron systems in
Sec.~\ref{sec:spectral_functions}.  Experimentally, the spectral functions
can in principle be studied by angle-resolved photoemission spectroscopy.
However, we are not aware of such experiments on one-dimensional
conductors in the regime of strong interactions.  A more promising
approach is to observe momentum-resolved tunneling between two parallel
quantum wires.  Such measurements have been recently reported by
Auslaender \emph{et al.}\cite{auslaender1,auslaender2} The regime of
strong interactions can be achieved in a quantum wire by reducing the
electron density.  Unfortunately, at low densities the effects of disorder
are also amplified, and the tunneling into states of given momentum is no
longer possible.\cite{auslaender2} Thus to observe the predicted features
in the spectral functions one would have to manufacture quantum wires with
even less disorder than in Refs.~\onlinecite{auslaender1,auslaender2}.

Our results can be compared with previous theoretical studies of the
spectral functions of one-dimensional electron systems.  In particular,
Meden and Sch\"onhammer\cite{meden} and Voit\cite{voit} studied the
spectral functions in the framework of the bosonization approach.  In the
case of a strongly interacting system their results are valid only at
energies below $J$.  However, they can still be compared with our results
at $\omega\gg J/\hbar$ by taking the formal limit $v_\sigma\to0$ in the
bosonization results, as suggested by Penc \textit{et al.}\cite{penc1} The
spectral functions of Refs.~\onlinecite{meden,voit} show singularities at
$k=k_F\pm\omega/v_\rho$, which are not sensitive to the spin velocity
$v_\sigma$.  As expected, the power-law behavior of the spectral functions
at those singularities is in agreement with our results
(\ref{eq:A_singularity}) and (\ref{eq:A_secondary_singularity}).
Similarly, in the $v_\sigma\to0$ limit, the singularities at
$\omega=\pm v_\sigma(k-k_F)$ of the spectral functions of
Refs.~\onlinecite{meden, voit} show the same power-law frequency
dependence as the Gaussian peak (\ref{eq:A^pm_result}) in the tail region
(at $k=k_F$).

Since the bosonization technique accounts only for the electrons near the
Fermi points, our results for $k$ away from $\pm k_F$ cannot be compared
with those of Refs.~\onlinecite{meden,voit}.  In particular, the main
contribution to the tunneling density of states at low energies is due to
the peak in the spectral functions centered at $k=0$,
Eq.~(\ref{eq:A^pm_result}).  Consequently, our expression (\ref{eq:nu})
for the density of states in the case of the Hubbard model ($K=1$) gives a
larger result $\nu\propto \varepsilon^{-1/2}$ than the Fermi-surface
contribution $\nu\propto \varepsilon^{-3/8}$ by Penc \textit{et
  al.}\cite{penc1} The inverse square-root dependence of the density of
states on energy was obtained earlier\cite{cheianov1,cheianov2,fiete1} in
the case of $T\gg J$.  Our results show\cite{brief} that the same
dependence holds also for $T\lesssim J$.  Physically the enhancement of
the density of states at low energy is analogous to that in the X-ray
absorption edge problem,\cite{mahan} with the spin excitations creating
the the effective core-hole potential for the holons.\cite{penc2,brief}

A detailed comparison can be made between our results for the spectral
function and those of Penc \textit{et al.}\cite{penc2} The latter work
studied numerically the quarter-filled Hubbard model in the limit of
infinite on-site repulsion $U$.  To compare their results with ours one
should assume $K=1$, and consequently $\zeta_0=\zeta_1=1/16$.  Our results
(\ref{eq:A_singularity}) and (\ref{eq:A_3kF_singularity}) indicate that
power-law peak with exponents $\zeta_{0,1}-1/2=-7/16$ should appear at
$k=k_F+\omega/v_\rho$ and $k=3k_F+\omega/v_\rho$.  The data of
Ref.~\onlinecite{penc2} does show singularities at those lines in the
$(k,\omega)$ plane.  In addition, as $\omega\to0$ the singularity at
$k=k_F+\omega/v_\rho$ is expected to grow as $\omega^{-7/16}$, whereas the
one at $k=3k_F+\omega/v_\rho$ is expected to be suppressed as $\omega
^{9/16}$.  The data of Ref.~\onlinecite{penc2} does show this qualitative
behavior.  Finally, the data\cite{penc2} clearly shows a weak feature at
$k=k_F-\omega/v_\rho$, which becomes more prominent at $\omega\to0$.  This
feature is consistent with our result (\ref{eq:A_secondary_singularity})
which at $\zeta_0=1/16$ behaves as
$\omega^{-15/16}[k-(k_F-\omega/v_\rho)]^{1/16}$.

One should note that the numerical data\cite{penc2} does not show a peak
at $k=0$ that we expect based on Eq.~(\ref{eq:A^pm_result}).  At
$\omega\to+0$ the spectral function appears to be very small.  This can be
understood as a result of smallness of $c^+_\gamma(\pi)\approx 0.044$.  At
$\omega\to-0$ the spectral function\cite{penc2} is not small, but instead
of a Gaussian peak it shows a rather flat minimum at $k=0$.  This can be
understood by noticing that the $k$-dependent prefactor
$c^-_\gamma(q_k)/\Gamma\big(\zeta(k)\big)$ in Eq.~(\ref{eq:A^pm_result})
has a minimum at $k=0$.  In the limit $|\omega|/D_\rho\to0$ the last
factor in Eq.~(\ref{eq:A^pm_result}) dominates, and we find a peak.
However, in a finite system one cannot access the values of
$\hbar|\omega|$ below the level spacing.  Substituting the parameters of
the Hubbard chain used in Ref.~\onlinecite{penc2} into
Eq.~(\ref{eq:A^pm_result}) and using the approximation
(\ref{eq:c^-muskhelishvili}) for $c^-_\gamma(q)$, we find
\begin{equation}
  \label{eq:peak}
  \ln A^-_\gamma(k,\omega) = {\rm const} + 
        \left(2.03-\frac{2}{\pi^2}\ln\frac{D_\rho}{\hbar|\omega|}\right) k^2
\end{equation}
at $k\to0$.  Thus the spectral function should have a peak if
$\ln(D_\rho/\hbar|\omega|)>10$.  On the other hand the finite level
spacing on the quarter-filled lattice of 228 sites\cite{penc2} limits the
frequencies such that
$\ln(D_\rho/\hbar|\omega|)\lesssim\ln(228/2\pi)\approx3.6$.  Thus to find
the peak at $k=0$ significantly longer systems should be studied.  In the
spin-incoherent regime $T\gg J$ a similar interplay of the Gaussian peak
in the spectral function with the minimum of the prefactor was discussed
in Ref.~\onlinecite{fiete3}.

\begin{acknowledgments}
  The authors are grateful to K. Penc for helpful discussions and to the
  Aspen Center for Physics, where part of this work was done, for
  hospitality.  This work was supported by the U. S. Department of Energy,
  Office of Science, under Contract No.~DE-AC02-06CH11357, by Grant-in-Aid
  for Scientific Research (Grant No.~16GS0219) from MEXT of Japan, and by
  NSF DMR Grants 0237296 and 0439026.
\end{acknowledgments}

\appendix

\section{Anticommutation of operators (\ref{eq:our_operators})}
\label{sec:anticommutation}

Since electrons are fermions, in one dimension their field operators are
expected to satisfy the following anticommutation relations:
\begin{subequations}
  \label{eq:electron_anticommutation_relations}
  \begin{eqnarray}
    \psi_\gamma(x)\psi_{\gamma'}(y)
      +\psi_{\gamma'}(y)\psi_\gamma(x)
    &=&0,
\label{eq:psi-psi}
\\
    \psi_\gamma^\dagger(x)\psi_{\gamma'}^\dagger(y)
      +\psi_{\gamma'}^\dagger(y)\psi_\gamma^\dagger(x)
    &=&0,
\label{eq:psidagger-psidagger}
\\
    \psi_\gamma(x)\psi_{\gamma'}^\dagger(y)
      +\psi_{\gamma'}^\dagger(y)\psi_\gamma(x)
    &=&\delta_{\gamma\gamma'}\delta(x-y).
\label{eq:psi-psidagger}
  \end{eqnarray}
\end{subequations}
Here we check that our form of electron operators (\ref{eq:our_operators})
is consistent with the relations
(\ref{eq:electron_anticommutation_relations}). 

We start by discussing the commutation relations of operators
$Z_{l,\gamma}$ and $Z_{l,\gamma}^\dagger$.  By definition these operators
act on the spin chain (\ref{eq:Heisenberg}) and change the number of sites
as follows.  Operator $Z_{l,\gamma}$ removes site $l$ from the spin chain
if that site has spin $\gamma$ and gives zero otherwise.  Conversely, the
operator $Z_{l,\gamma}^\dagger$ adds a new site $l$ with spin $\gamma$ to
the spin chain by inserting it between the sites $l-1$ and $l$.  

Let us consider the effect of operator $Z_{l,\gamma}Z_{l',\gamma'}$ with
$l<l'$.  This operator first removes spin $\gamma'$ from site $l'$ and then
spin $\gamma$ from site $l$.  Alternatively, one can first remove site $l$
and notice that the numbering of all sites after $l$ has shifted by 1.
Thus to achieve the same result, at the second step one needs to remove
site $l'-1$.  We therefore conclude
\begin{subequations}
\begin{eqnarray}
  \label{eq:ZZ}
  Z_{l,\gamma}Z_{l',\gamma'}=Z_{l'-1,\gamma'}Z_{l,\gamma},
\quad
  l<l'.
\end{eqnarray}
Repeating these arguments with operators $Z^\dagger$, we find
\begin{eqnarray}
  \label{eq:ZdaggerZdagger}
  Z_{l,\gamma}^\dagger Z^\dagger_{l',\gamma'}
  =Z_{l'+1,\gamma'}^\dagger Z_{l,\gamma}^\dagger,
\quad
  l\leq l',
\\
  \label{eq:ZZdagger}
  Z_{l,\gamma}Z_{l',\gamma'}^\dagger=Z_{l'-1,\gamma'}^\dagger Z_{l,\gamma},
\quad
  l<l'.
\end{eqnarray}
\end{subequations}

Let us check Eq.~(\ref{eq:psi-psidagger}) at $x<y$.  We start by writing
the first term as
\begin{equation}
  \psi_\gamma(x)\psi_{\gamma'}^\dagger(y)
  = \Psi(x)Z_{l(x),\gamma}Z_{l(y),\gamma'}^\dagger\Psi^\dagger(y)
\label{eq:first-term}
\end{equation}
and noticing that according to the definition (\ref{eq:l(x)}) the holon
created at point $y$ is counted in $l(y)$, but not in $l(x)$.  Thus
$l(x)<l(y)$ and one can use Eq.~(\ref{eq:ZZdagger}),
\[
  \psi_\gamma(x)\psi_{\gamma'}^\dagger(y)
  = \Psi(x)Z_{l(y)-1,\gamma'}^\dagger Z_{l(x),\gamma}\Psi^\dagger(y).
\]
Using the definition (\ref{eq:l(x)}) again, we find that at $x<y$ the
operators $Z_{l(x),\gamma}$ and $\Psi^\dagger(y)$ commute, while
$\Psi(x)Z_{l(y)-1,\gamma'}^\dagger = Z_{l(y),\gamma'}^\dagger\Psi(x)$.
Thus we conclude
\begin{eqnarray*}
  \psi_\gamma(x)\psi_{\gamma'}^\dagger(y)
  &=& Z_{l(y),\gamma'}^\dagger\Psi(x)\Psi^\dagger(y) Z_{l(x),\gamma}
\\
  &=& -Z_{l(y),\gamma'}^\dagger\Psi^\dagger(y)\Psi(x) Z_{l(x),\gamma}
\\
  &=& - \psi_{\gamma'}^\dagger(y)\psi_\gamma(x),
\end{eqnarray*}
in agreement with Eq.~(\ref{eq:psi-psidagger}).  One can easily perform a
similar check of Eqs.~(\ref{eq:psi-psi}), (\ref{eq:psidagger-psidagger}),
and the case of $x>y$.

At $x=y$ the relations (\ref{eq:psi-psi}) and
(\ref{eq:psidagger-psidagger}) for the operators (\ref{eq:our_operators})
are trivially satisfied, because $\Psi(x)\Psi(x)=
\Psi^\dagger(x)\Psi^\dagger(x)=0$.  On the other hand, the relation
(\ref{eq:psi-psidagger}) is less straightforward.  From
Eq.~(\ref{eq:first-term}) at $x=y$ we get
\begin{equation}
  \psi_\gamma(x)\psi_{\gamma'}^\dagger(x)
  = \Psi(x)\Psi^\dagger(x)\delta_{\gamma\gamma'},
\label{eq:first-term_x=y}
\end{equation}
because the operator $Z_{l,\gamma}Z_{l,\gamma'}^\dagger$ first creates a
site with spin $\gamma'$ and then removes the same site with spin
$\gamma$.  The second term in the left-hand side of
Eq.~(\ref{eq:psi-psidagger}) becomes
\begin{equation}
  \psi_{\gamma'}^\dagger(x)\psi_\gamma(x)
  = \Psi^\dagger(x)\Psi(x)Z_{l(x),\gamma'}^\dagger Z_{l(x),\gamma}^{}.
\label{eq:second-term_x=y}
\end{equation}
Contrary to the expectation based upon Eq.~(\ref{eq:psi-psidagger}),
$Z_{l,\gamma'}^\dagger Z_{l,\gamma}^{}\neq \delta_{\gamma\gamma'}$.

The reason for this apparent discrepancy is that our operators
(\ref{eq:our_operators}) act in a restricted Hilbert space, where two
electrons cannot occupy the same point $x$, even if their spins are
opposite.  This is a fundamental feature of our theory, which reflects
the fact that electrons repel each other very strongly.  In this
restricted space, the operator
\begin{equation}
  \label{eq:left-hand-side}
  \psi_\gamma(x)\psi_{\gamma'}^\dagger(x)
   +\psi_{\gamma'}^\dagger(x)\psi_\gamma(x)
\end{equation}
is \emph{not\/} equivalent to $\delta_{\gamma\gamma'}$.  [Here it is
convenient to view $x$ as a discrete coordinate and replace
$\delta(x-x)\to 1$ in Eq.~(\ref{eq:psi-psidagger})].  Indeed, in our
Hilbert space the state at point $x$ can be either empty, or occupied with
a single electron, with possible spins $\uparrow$ or $\downarrow$.  When
acting on these states, the operator (\ref{eq:left-hand-side}) has the
following effect:
\begin{eqnarray*}
  |0\rangle &\to& \delta_{\gamma\gamma'},
\\
  |\!\uparrow\rangle &\to& \delta_{\gamma,\uparrow} |\gamma'\rangle,
\\
  |\!\downarrow\rangle &\to& \delta_{\gamma,\downarrow} |\gamma'\rangle.
\end{eqnarray*}
It is easy to check that the sum of operators (\ref{eq:first-term_x=y})
and (\ref{eq:second-term_x=y}) has exactly the same effect on these three
states.  Thus our expressions (\ref{eq:our_operators}) for the electron
operators have correct anticommutation relations.

\section{Bosonization of the operators $Z_{l,\gamma}$ in the case of
  non-vanishing magnetization}
\label{sec:magnetization}

In Sec.~\ref{sec:Z-bosonization} we bosonized the operators $Z_{l,\gamma}$
in the SU(2)-symmetric case, when no magnetic field is applied to the
system.  In the presence of the field, the ground state of the system has
unequal densities of electrons with spins $\uparrow$ and $\downarrow$,
giving rise to a finite magnetization $m=2\langle S_l^z\rangle$.  As a
result the Fermi sea of the Jordan-Wigner fermions
(Fig.~\ref{fig:groundstate}) expands to accommodate their increased density
$\langle a_l^\dagger a_l^{}\rangle=(1+m)/2$, see
Eq.~(\ref{eq:Jordan-Wigner}).  The Fermi points corresponding to this
density are
\begin{equation}
  \label{eq:Fermi_points_magnetization}
    q_L=\frac{\pi}{2}(1-m),
\quad
  q_R=\frac{\pi}{2}(3+m).
\end{equation}
Our bosonization procedure in Sec.~\ref{sec:Z-bosonization} was performed
for $m=0$, but the derivation can be easily generalized to the case of
$m>0$ by using the proper values (\ref{eq:Fermi_points_magnetization}) of
$q_{L,R}$ instead of Eq.~(\ref{eq:Fermi_points}).  The resulting
bosonization expressions for the operators $Z_{l,\gamma}$ are given by
\begin{eqnarray}
    Z_{l,\gamma}
     &=&e^{i\frac{\pi}{2}(1\mp m) l}
        e^{\mp i[\varphi(l)+\frac{1}{2}(1\mp m)\vartheta(l)]}
\nonumber\\
     &&+e^{-i\frac{\pi}{2}(1\mp m) l}
        e^{\pm i[\varphi(l)-\frac{1}{2}(1\mp m)\vartheta(l)]},
\label{eq:Z_bosonization_magnetization}
\end{eqnarray}
with the upper and lower signs corresponding to $\gamma=\,\uparrow$ and
$\downarrow$, respectively.

It is instructive to substitute
Eq.~(\ref{eq:Z_bosonization_magnetization}) into our expression
(\ref{eq:our_annihilation_operator_bosonized}) for the electron
annihilation operators and compare the resulting bosonization formulas
with the standard expressions (\ref{eq:electron_bosonization}).  As we
discussed in Sec.~\ref{sec:two-step}, the right-moving electron is
constructed out of a right-moving holon and a left-moving spinon.  We
therefore combine the second term in
Eq.~(\ref{eq:Z_bosonization_magnetization}) with the first term in the
parentheses in Eq.~(\ref{eq:our_annihilation_operator_bosonized}) and
obtain
\begin{equation}
  \psi_{R\gamma}(x)
    =\frac{e^{-i\theta(x)}}{\sqrt{2\pi\alpha}}\, 
      e^{i\frac{1\pm m}{2}[k_F^h x+\phi(x)]}
      e^{\pm i[\varphi(nx)-\frac{1\mp m}{2}\vartheta(nx)]}.
  \label{eq:right-mover}
\end{equation}
This result should be compared with the standard bosonization expression
(\ref{eq:electron_bosonization_right}).  

As expected, instead of
$k_F=k_F^h/2$ we find that the Fermi momentum is now a function of the
magnetization:
\begin{equation}
  \label{eq:k_F_magnetization}
  k_{F\uparrow}=\frac{1+m}{2}k_F^h,
\quad
  k_{F\downarrow}=\frac{1-m}{2}k_F^h.
\end{equation}
In addition, by comparing Eq.~(\ref{eq:right-mover}) and the analogous
expression for the left-moving electron
\begin{equation}
  \psi_{L\gamma}(x)
    =\frac{e^{-i\theta(x)}}{\sqrt{2\pi\alpha}}\, 
      e^{-i\frac{1\pm m}{2}[k_F^h x+\phi(x)]}
      e^{\mp i[\varphi(nx)+\frac{1\mp m}{2}\vartheta(nx)]}
  \label{eq:left-mover}
\end{equation}
with Eq.~(\ref{eq:electron_bosonization}) one finds the following
relations between the bosonic fields:
\begin{eqnarray}
  \label{eq:bosonic_relations}
  \phi(x) &=& \sqrt2\, \phi_\rho(x),
\\
  \theta(x) &=& \frac{\theta_\rho(x) + m\, \theta_\sigma(x)}{\sqrt2},
\\
  \varphi(l)&=& \frac{\phi_\sigma(l/n) - m\, \phi_\rho(l/n)}{\sqrt2},
\\
  \vartheta(l)&=& \sqrt2\, \theta_\sigma(l/n).
\end{eqnarray}
These relations generalize our earlier expressions
(\ref{eq:relations_between_charge_variables}a,b) and
(\ref{eq:relations_between_spin_variables}a,b) to the case of
non-vanishing magnetization.  It is worth noting that at $m>0$ the
original charge and spin boson modes are mixed.  This mixing was discussed
in Ref.~\onlinecite{hikihara}.

\section{Solution of equation (\ref{eq:minus-to-plus}) with conditions
  (\ref{eq:spinon_Fermi_surface})}
\label{sec:muskhelishvili}

In this Appendix we show that the solution of Eq.~(\ref{eq:minus-to-plus})
with conditions (\ref{eq:spinon_Fermi_surface}) has the form of
Eqs.~(\ref{eq:c^+muskhelishvili}) and (\ref{eq:c^-muskhelishvili}).  We
first rewrite the integral equation (\ref{eq:minus-to-plus}) in terms of
complex variables $w=e^{iq}$ and $z=e^{iq'}$.  Using the fact that
$c^+_\gamma$ is even function of $q$, we find
\begin{equation}
  \label{eq:minus-to-plus_complex}
  c^-(w)=\frac{w^2+1}{2w}\,c^+(w)
         -\frac{w}{2\pi i}
          \oint\frac{dz}{z-w}\,\frac{z^2-1}{z^2}\,c^+(z).
\end{equation}
Here we use the notations $c^\pm_\gamma(q)=c^\pm(w)$; the integral is
taken over the unit circle $|z|=1$ in counterclockwise direction.

According to the condition (\ref{eq:spinon_Fermi_surface_minus}) function
$c^-(w)$ vanishes when $w$ is on the right semicircle $R$ (defined as
$w=e^{iq}$ with $q$ between $-\pi/2$ and $\pi/2$).  In additions, $c^+(z)$
vanishes on the left semicircle, see
Eq.~(\ref{eq:spinon_Fermi_surface_plus}).  Thus for $w\in R$ we have
\begin{equation}
  \label{eq:c^-_integral_equation}
  \frac{w^2+1}{w^2}\,c^+(w)
   = \frac{1}{\pi i}
     \int_R\frac{dz}{z-w}\,\frac{z^2-1}{z^2}\,c^+(z),
  \quad
  w\in R.
\end{equation}
Here we assume that the contour $R$ is traversed in the counterclockwise
direction, from $z=-i$ to $z=i$.

This equation can be solved using the theory of singular integral
equations.\cite{muskhelishvili}  It will be convenient to introduce a new
unknown function
\begin{equation}
  \label{eq:phi}
  \phi(z)= \frac{z^2-1}{z^2}\, c^+(z).
\end{equation}
Then the integral equation (\ref{eq:c^-_integral_equation}) takes the form
\begin{equation}
  \label{eq:integral_equation}
  \frac{w^2+1}{w^2-1}\, \phi(w) 
         =\frac{1}{\pi i}\int_R dz\, \frac{\phi(z)}{z-w},
  \quad w\in R.
\end{equation}
In solving this equation we will assume that the unknown function
$\phi(w)$ is analytic with possible exception of an integrable
singularities at the ends $z=\pm i$ of the contour $R$.  Let us introduce
a new function
\begin{equation}
  \label{eq:Phi}
  \Phi(w)=\frac{1}{2\pi i}\int_R dz\, \frac{\phi(z)}{z-w}.
\end{equation}
Obviously $\Phi(w)$ is analytic everywhere except the contour $R$ and
approaches zero at infinity.  At the contour $\Phi(w)$ has a branch cut.
When $w$ approaches $R$ from the left or right, the function $\Phi(w)$
takes the values
\begin{equation}
  \label{eq:Phi+-}
  \Phi^\pm(w)=\frac{1}{2\pi i}\int_R dz\, \frac{\phi(z)}{z-w\pm \delta},
\quad
  \delta\to+0.
\end{equation}
One can easily see that for  $w \in R$ we have
\begin{eqnarray}
  \Phi^+(w)-\Phi^-(w)&=&\phi(w),
  \label{eq:Phi_difference}
\\
  \Phi^+(w)+\Phi^-(w)&=&\frac{1}{\pi i}\int_R dz\, \frac{\phi(z)}{z-w}.
\end{eqnarray}
Substituting these relations in Eq.~(\ref{eq:integral_equation}), we find
that the values of $\phi(w)$ on the two sides of the branch cut satisfy
the following linear relation:
\begin{equation}
  \label{eq:Phi_relations}
  \Phi^+(w)=w^2\Phi^-(w),
  \quad
  w\in R.
\end{equation}
An analytic function that satisfies these conditions on the two sides of
the contour $R$, falls off to zero at infinity, and does not diverge
faster than $1/(w\pm i)$ at the ends of the contour, is unique up to an
arbitrary numerical coefficient.  It can be found using the techniques
discussed in \S79 of Ref.~\onlinecite{muskhelishvili}.   The
solution is
\begin{equation}
  \label{eq:Hilbert}
  \Phi(w)=\frac{1}{w^2+1}
           \exp\left(
               \frac{1}{\pi i}
               \int_R\frac{\ln z}{z-w}\,dz
               \right),
\end{equation}
where the logarithm is defined with the branch cut along the negative real
axis.  It is easy to check directly that the function (\ref{eq:Hilbert})
does satisfy the above conditions.

For $w\in R$, by combining Eqs.~(\ref{eq:phi}), (\ref{eq:Phi_difference}),
and (\ref{eq:Phi_relations}) one finds $c^+(w)=\Phi^+(w)$.  Then,
substituting $w=e^{iq}$ one finds the result (\ref{eq:c^+muskhelishvili}).

Our next goal is to find $c^-(w)$ for $w$ on the left unit semicircle $L$,
defined by $w=e^{iq}$ with $q$ between $\pi/2$ and $3\pi/2$.  Using
Eq.~(\ref{eq:minus-to-plus_complex}) and noticing that $c^+(w)=0$ on $L$,
we find
\begin{equation}
  \label{eq:c^-_complex}
  c^-(w)= -\frac{w}{2\pi i}
          \int_R\frac{dz}{z-w}\,\frac{z^2-1}{z^2}\,c^+(z)
        =-w\Phi(w),
\end{equation}
see Eqs.~(\ref{eq:phi}) and (\ref{eq:Phi}).  Then, substituting
$w=e^{iq}$, we obtain the result (\ref{eq:c^-muskhelishvili}) with the
same normalization constant as in Eq.~(\ref{eq:c^+muskhelishvili}).

\section{Mapping of the bosonized Heisenberg spin chain to the spin 
sector of weakly interacting electron system}
\label{sec:g-ology}

\subsection{Consequences of the spin-rotation symmetry}
\label{sec:SU2}

In Sec.~\ref{sec:c-correction} we have evaluated the first correction
(\ref{eq:delta_c^+bosonization_result}) to the correlator $c^+_\gamma(q)$
above the Fermi point $q=\pi/2$, where the simple bosonization result
(\ref{eq:c^+bosonization_result}) vanishes.  The correction originated
from two sources.  First, we found the contribution due to the deviation
of the quadratic part of the Hamiltonian from the fixed point, $\mathcal
K\neq1/2$, and then we included the sine-Gordon term (\ref{eq:V}).  The
latter correction turned out to be larger than the former one by a factor
of two.  Here we show that this is a result of the spin-rotation symmetry
of the problem.

To this end we utilize the equivalence of the bosonized Hamiltonians of
the Heisenberg spin chain (\ref{eq:Heisenberg_bosonized}) and the spin
part (\ref{eq:H_sigma}) of the Hamiltonian of weakly interacting
electrons.  The exact form of the electron-electron interactions does not
affect the general form of the Hamiltonian
(\ref{eq:standard_bosonization}).  In the simplest case, one can consider
only backscattering of electrons by each other.  For the electrons in the
vicinity of the Fermi level the most general form of backscattering
Hamiltonian is
\begin{equation}
  \label{eq:backscattering_interaction}
  g_{\gamma \gamma' \delta \delta'} 
   \psi_{L\gamma}^\dagger\psi_{R\delta}^\dagger
   \psi^{}_{L\gamma'}\psi^{}_{R\delta'},
\end{equation}
with the coupling constant
\begin{equation}
  g_{\gamma \gamma' \delta \delta'}= \frac12
  g_{1} (\delta_{\gamma\gamma'}\delta_{\delta\delta'}
  + {\bm\sigma}_{\gamma\gamma'}^{}\cdot{\bm\sigma}_{\delta\delta'}^{}),
  \label{eq:g1_symmetric}
\end{equation}
where ${\bm\sigma}_{\gamma\gamma'}$ is the vector of standard Pauli
matrices and $g_1$ is the $2k_F$-Fourier component of the interaction
potential. In Eq.~(\ref{eq:backscattering_interaction}) the summation over
repeating spin indices is implied.

The exact form of the coupling of electron spins in the Hamiltonian
(\ref{eq:backscattering_interaction}), (\ref{eq:g1_symmetric}) is dictated
by the SU(2) symmetry of the problem with respect to the rotation of
electron spins.  Ignoring this symmetry for the moment, we will view
Eq.~(\ref{eq:g1_symmetric}) as a special case of tensor
\begin{eqnarray}
  g_{\gamma \gamma' \delta \delta'}&=& \frac12
  (g_{1\rho} \delta_{\gamma\gamma'}\delta_{\delta\delta'}
  +g_{1z} \sigma^z_{\gamma\gamma'}\sigma^z_{\delta\delta'}
  \nonumber\\
  &&+g_{1x} \sigma^x_{\gamma\gamma'}\sigma^x_{\delta\delta'}
    +g_{1y} \sigma^y_{\gamma\gamma'}\sigma^y_{\delta\delta'}).
  \label{eq:g1}
\end{eqnarray}
Unlike Eq.~(\ref{eq:g1_symmetric}), this form of coupling violates the
spin-rotation symmetry, unless $g_{1x}=g_{1y}=g_{1z}$.  The standard
treatments of weakly-interacting electrons systems, including the
derivation of the bosonized Hamiltonian (\ref{eq:standard_bosonization}),
start with two constants, $g_{1\parallel}=g_{1z}$ and $g_{1\perp}=g_{1x} =
g_{1y}$, and eventually equate $g_{1\parallel}$ and $g_{1\perp}$.

Interaction constants $g_{1\rho}$, $g_{1\parallel}$ and $g_{1\perp}$
affect different terms of the bosonized Hamiltonian
(\ref{eq:standard_bosonization}).  Parameter $g_{1\rho}$ corresponds to
density-density coupling and affects the Hamiltonian of the charge degrees
of freedom via renormalization of $v_\rho$ and $K_\rho$.  It does not
affect $H_\sigma$ and for our purposes can be ignored.  The coupling
constant $g_{1\parallel}$ enters via $K_\sigma=1+g_{1\parallel}/2$,
whereas the spin flip scattering accounted for by $g_{1\perp}$ transforms
to the sine-Gordon term in Eq.~(\ref{eq:H_sigma}).

Our calculation of the correction (\ref{eq:delta_c^+bosonization_result})
neglected the sine-Gordon term.  Thus $g^2$ in
Eq.~(\ref{eq:delta_c^+bosonization_result}) is in fact $g_{1z}^2$.  Since
$c^+_\gamma(q)$ is invariant with respect to spin rotations, identical
contributions should be expected from coupling constants $g_{1x}$ and
$g_{1y}$.  While evaluating the correction to $c^+_\gamma(q)$ due to the
sine-Gordon term (\ref{eq:V}), we accounted for both $g_{1x}$ and $g_{1y}$
and, as expected, obtained twice the result
(\ref{eq:delta_c^+bosonization_result}).

\subsection{Alternative evaluation of the correlators $c^+_\gamma(q)$ near
  the Fermi points}

\label{sec:alternative}

The mapping of the bosonized Hamiltonian (\ref{eq:Heisenberg_bosonized})
of the Heisenberg spin chain and the Hamiltonian $H_\sigma$,
Eq.~(\ref{eq:H_sigma}), describing the dynamics of the spin sector of
weakly interacting electron gas, enables one to obtain an alternative
expression for the correlators $c^\pm_\gamma(q)$ near the Fermi point.
Let us consider the Green's function of right-moving electrons $G_R(x,t)$
traced over the spin indices.  It is well known\cite{dzyaloshinskii} that
asymptotically at large $x$ and $t$ it separates into a product of charge
and spin factors,
\begin{equation}
  \label{eq:spin_charge_separation}
  G_R(x,t)=\frac{1}{\pi} g_\rho(x,t)g_\sigma(x,t).
\end{equation}
The two factors are most easily computed using the bosonization
transformation (\ref{eq:electron_bosonization}), in which case the charge
and spin factors are obtained by averaging the exponentials of bosonic
fields $\phi_\rho$, $\theta_\rho$ and $\phi_\sigma$, $\theta_\sigma$ in
Eq.~(\ref{eq:electron_bosonization_right}), respectively.  In the absence
of electron-electron interactions the parameters of the Hamiltonian
(\ref{eq:standard_bosonization}) take unperturbed values
$v_\rho=v_\sigma=v_F$, $K_\rho=K_\sigma=1$, and $g_{1\perp}=0$.  Then one
finds
\begin{equation}
  \label{eq:greens_functions_unperturbed}
  g_\rho^{(0)}(x,t)=g_\sigma^{(0)}(x,t)=
       \frac{1}{(x-v_F t + i\delta{\,\rm sgn\,}t)^{1/2}},
\end{equation}
and Eq.~(\ref{eq:spin_charge_separation}) reproduces the standard
expression for the Green's function of non-interacting electrons.

In the presence of interactions the parameters of the Hamiltonian
(\ref{eq:standard_bosonization}) renormalize, and the charge and spin
components of the Green's function show non-trivial behavior.  We showed
in Sec.~\ref{sec:two-step} that the exponentials of the bosonic fields
$\phi_\sigma$ and $\theta_\sigma$ in
Eq.~(\ref{eq:electron_bosonization_right}) are equivalent to the bosonized
expression (\ref{eq:Z_bosonization}) for the operators $Z_{l,\gamma}$.
Thus the correlator $\langle Z_{l,\gamma}Z_{0,\gamma}^\dagger\rangle$ in
the definition (\ref{eq:c^+}) of $c^+_\gamma(q)$ can be found from
$g_\sigma(x,t)$ at $x=l/n$ and $t=0$.

We will find $g_\sigma(x,t)$ by calculating electron Green's function
(\ref{eq:spin_charge_separation}) and identifying its spin component.
Since the Hamiltonian (\ref{eq:H_sigma}) of the spin sector is universal
at low energies, the specific form of the electron-electron interactions
is not important.  It is most convenient to choose the form
(\ref{eq:backscattering_interaction}) with the coupling constant
(\ref{eq:g1}) chosen so that $g_{1\rho}=0$ and $g_{1x}=g_{1y}=g_{1z}=g_1$.
In this case the interactions do not affect the charge sector, and
$g_\rho(x,t)$ retains its unperturbed value
(\ref{eq:greens_functions_unperturbed}), and 
\begin{equation}
  \label{eq:g_sigma_from_G}
  g_\sigma(x,t)=\pi(x-v_F t + i\delta{\,\rm sgn\,}t)^{1/2} G_R(x,t).
\end{equation}
The electronic Green's function can be studied using straightforward
perturbation theory in the coupling constant $g_1$.

\begin{figure}[t]
 \resizebox{.42\textwidth}{!}{\includegraphics{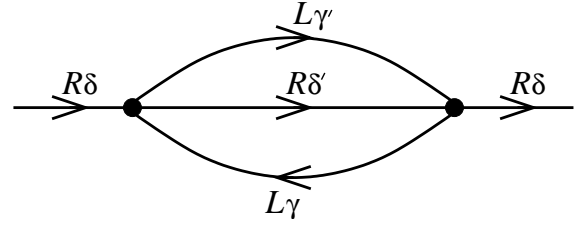}}
\caption{\label{fig:2nd-order} The second-order diagram for the perturbative
calculation of the electronic Green's function $G_R(x,t)$.}
\end{figure}

The first non-vanishing contribution to $G_R(x,t)$ appears in the second
order and is given by the diagram Fig.~\ref{fig:2nd-order}.  The
respective correction to the Green's function is
\begin{equation}
  \label{eq:diagram}
  \delta G_{R}(k,\omega) 
    = 2 G_R^{(0)}(k,\omega)\Sigma(k,\omega)G_R^{(0)}(k,\omega),
\end{equation}
where the factor of 2 accounts for the trace over spin variables and the
self-energy
\begin{eqnarray}
  \Sigma(k,\omega) &=& \frac{3g_1^2}{2}
                       \int\frac{dk_1 dk_2 d\omega_1 d\omega_2}{(2\pi)^4}
                       G_L^{(0)}(k_1,\omega_1)  G_L^{(0)}(k_2,\omega_2) 
\nonumber\\
 &&\times           G_R^{(0)}(k+k_1-k_2,\omega+\omega_1-\omega_2).
  \label{eq:Sigma}
\end{eqnarray}
Here the factor $3g_1^2$ appears as the sum $g_{1x}^2+g_{1y}^2+g_{1z}^2$.
Substituting the unperturbed Green's functions
\begin{subequations}
    \label{eq:Greens_functions_unperturbed}
\begin{eqnarray}
  G_L^{(0)}(k,\omega) &=& \frac{1}{\omega+v_F k - i\delta\, {\rm sgn}\,k},
\\
  G_R^{(0)}(k,\omega) &=& \frac{1}{\omega-v_F k + i\delta\, {\rm sgn}\,k},
\end{eqnarray}
\end{subequations}
and performing integration with respect to $\omega_1$, $\omega_2$, and one
of the momenta, we find
\begin{equation}
  \label{eq:Sigma_final}
  \Sigma(k,\omega) = \Sigma_v(k) + \widetilde\Sigma(k,\omega),
\end{equation}
where
\begin{eqnarray}
  \label{eq:Sigma_v}
  \Sigma_v(k)&=&-\frac{3g_1^2 k}{16\pi^2v_F},\\
  \widetilde\Sigma(k,\omega) &=& \frac{3g_1^2(\omega-v_F k)}{16\pi^2v_F}
          \int_0^\infty
       \bigg(
          \frac{\theta(k+q)}{\omega-v_F k - 2v_F q + i\delta}
\nonumber\\
       &&
          -\frac{\theta(-k+q)}{\omega-v_F k + 2v_F q - i\delta}
       \bigg)e^{-\alpha q}\,dq.
       \label{eq:tilde_Sigma}
\end{eqnarray}
Here we have introduced the short-distance cutoff $\alpha$ for the
electron-electron interactions.  In principle, this cutoff may not
coincide with the bandwidth cutoff $\alpha$ used in the bosonization
procedure.  This distinction is not important for the present discussion.

The two second-order contributions to the electron self-energy have very
different meanings.  The term $\Sigma_v$ accounts for a small correction
to the velocity of spin excitations, which for our purposes can be
ignored.  On the other hand, $\widetilde\Sigma$ leads to the logarithmic
renormalization of the electron Green's function, which affects the
singular behavior of the correlators $c_\gamma^\pm (q)$ near the Fermi
points.  We therefore explore this correction in more detail.

\subsubsection{Logarithmic correction to the Green's function}
\label{sec:Greens_functions_renormalizations}

Let us now substitute the expression (\ref{eq:tilde_Sigma}) for the
self-energy $\Sigma(k,\omega)$ in Eq.~(\ref{eq:diagram}) and perform the
Fourier transformation to $x$ and $t$ variables.  The resulting correction
to the Green's function has the form
\begin{eqnarray}
  \delta G_{R}(x,t) &=&
\frac{3g_1^2}{32\pi^3 v_F^2}\,\frac{1}{x-v_F t +i\delta\,{\rm sgn\,}t}
\nonumber\\
&&\times
       \ln\frac{\alpha^2}{(\alpha+iv_F |t|)^2+x^2}.
  \label{eq:deltaG(x,t)}
\end{eqnarray}
This expression is consistent with the logarithmic renormalization of the
electron Green's function studied earlier in the $g$-ology theory, cf. Eq.
(4.24) of Ref.~\onlinecite{solyom}.  (Our Eq.~(\ref{eq:deltaG(x,t)}) is
obtained by neglecting the constant $g_2$ and replacing $4g_1^2$ with
$3g_1^2$ to account for the fact that we assume $g_{1\rho}=0$.)

We can now separate the spin component $g_\sigma(x,t)$ of the electron Green's
function $G_R(x,t)$ following the prescription (\ref{eq:g_sigma_from_G})
and obtain
\begin{eqnarray}
\hspace{-2.5em}
  g_\sigma(x,t)&=&\frac{1}{(x-v_F t + i\delta{\,\rm sgn\,}t)^{1/2}}
\nonumber\\
  &&\times
    \left(
       1-\frac{3y_1^2}{32}\,
         \ln\frac{(\alpha+iv_F |t|)^2+x^2}{\alpha^2}
    \right),
  \label{eq:g_sigma_log_correction}
\end{eqnarray}
where $y_1=g_1/\pi v_F$.  As expected, $g_s(x,+0)$ reproduces the
logarithmic correction to the correlator $\langle Z_{l\gamma}
Z^\dagger_{0,\gamma}\rangle$ used in
Eq.~(\ref{eq:delta_c^+bosonization_v1}), with the additional factor of 3
correctly included.

\subsubsection{Renormalization of the Green's function}
\label{sec:multiplicative_RG}

The logarithmic correction (\ref{eq:g_sigma_log_correction}) to the
Green's function grows at long distances, and can in principle become
large despite the smallness of the prefactor $y_1^2$.  To find out whether
this is the case, one can compute the Green's function in the leading
logarithm approximation.  We accomplish this by adopting the
multiplicative renormalization procedure of Ref.~\onlinecite{solyom}.  We
present the spin component of the Green's function as
\begin{equation}
  \label{eq:Greens_function_scaling}
  g_\sigma(x,t)=d(\xi)\,g_\sigma^{(0)}(x,t).
\end{equation}
Here the new function $d(\xi)$ is expected to depend on $x$ and $t$ very
slowly, via their logarithm,
\begin{equation}
  \label{eq:xi}
  \xi(x,t)=\frac12\ln\frac{(\alpha+iv_F |t|)^2+x^2}{\alpha^2}
\end{equation}
In particular, our result
(\ref{eq:g_sigma_log_correction}) has the form
(\ref{eq:Greens_function_scaling}) with
\begin{equation}
  \label{eq:d_perturbative}
  d(\xi)=1-\frac{3y_1^2}{16}\,\xi.
\end{equation}
As we increase $\xi$, the correction (\ref{eq:d_perturbative}) grows and
may no longer remain small.  In addition, the coupling constant $y_1$
itself depends on $\xi$ as
\begin{equation}
  \label{eq:y_1_renormalization}
  y_1(\xi) = \frac{y_1}{1+y_1\xi}.
\end{equation}
see Ref.~\onlinecite{solyom}.

Following the general prescription\cite{solyom} to account for the
multiplicative corrections to the Green's function, we use
Eq.~(\ref{eq:d_perturbative}) to write the renormalization group equation
upon $\ln d$, 
\begin{equation}
  \label{eq:multiplicative_RG}
  \frac{d}{d\xi}\ln d(\xi)=-\frac{3}{16}y_1^2(\xi).
\end{equation}
Solution of this equation with $y_1(\xi)$ given by
Eq.~(\ref{eq:y_1_renormalization}) has the form
\begin{equation}
  \label{eq:d_result}
  d(\xi)=\exp\left(-\frac{3}{16}\,\frac{y_1^2\xi}{1+y_1\xi}\right).
\end{equation}
At small $y_1\xi$ it reproduces the perturbative expansion
(\ref{eq:d_perturbative}), while in the limit $\xi\to\infty$ we obtain a
finite renormalization of the Green's function,
\begin{equation}
  \label{eq:d_infinity}
  g_\sigma(x,t) = \exp\left(-\frac{3y_1}{16}\right)
                 \frac{1}{(x-v_F t + i\delta{\,\rm sgn\,}t)^{1/2}}.
\end{equation}

In a weakly interacting electron gas the coupling constant $y_1\ll1$, and
the renormalization (\ref{eq:d_infinity}) can be ignored.  As interactions
become stronger, $y_1$ increases and reaches values of order unity.  Thus
one can expect that the coefficient $\chi$ in the asymptotes
(\ref{eq:c_sine-Gordon}) of the correlators $c^\pm_\gamma(q)$ near the
Fermi point will slowly decrease from its numerically obtained value
$\chi\sim0.8$ to a significantly lower number at $q\to\pi/2$.


\begin{thebibliography}{99}
\bibitem{footnote1} Similar physics results in the conclusion that the spin
  excitations of a one-dimensional electron system may carry electric
  charge.\cite{brazovskii,nayak}
\bibitem{footnote2} The same correlation function in coordinate
  representation was first introduced and studied numerically by exact
  diagonalization of spin chains of 26 sites by Ogata and
  Shiba.\cite{ogata} (Our notation $\langle Z^\dagger_{l\sigma}
  Z^{}_{0\sigma}\rangle$ corresponds to $2\omega_{0\to l}$ in
  Ref.~\onlinecite{ogata}.)
\end{thebibliography}
\end{document}